\documentclass[aps,prc,twocolumn,nofootinbib,floatfix,superscriptaddress]{revtex4-2}

\usepackage{graphicx}% Include figure files
\usepackage{dcolumn}% Align table columns on decimal point
\usepackage{amsmath}
\usepackage{amssymb}
\usepackage{psfrag}
\usepackage{epsfig}
\usepackage{epsf}
\usepackage{float}
\usepackage{slashed}
\usepackage{wasysym}

\usepackage{color}

\usepackage{multirow}

\newcommand{\nmax}{\ensuremath{N_{\rm max}}}
\newcommand{\hw}{\ensuremath{\hbar\omega}}
\newcommand{\fet}[1]{\mbox{\boldmath $#1$}}

\newcommand{\beq}{\begin{equation}}
\newcommand{\eeq}{\end{equation}}
\newcommand{\beqa}{\begin{eqnarray}}
\newcommand{\cbar}[0]{\bar c}
\newcommand{\eeqa}{\end{eqnarray}}
\newcommand{\nn}{\nonumber \\ }

\begin{document}

%\preprint{}

\title{Light nuclei with semilocal momentum-space regularized chiral interactions up to third order}


\author{P.~Maris}
\email[]{pmaris@iastate.edu}
\affiliation{Department of Physics and Astronomy, Iowa State
  University, Ames, Iowa 50011, USA}
\author{E.~Epelbaum}
\affiliation{Ruhr-Universit\"at
  Bochum, Fakult\"at f\"ur Physik und Astronomie, Institut f\"ur Theoretische Physik II, D-44780 Bochum, Germany}

\author{R.J.~Furnstahl}
\affiliation{Department of Physics, The Ohio State University, 
Columbus, Ohio 43210, USA}

\author{J.~Golak}
\affiliation{M.~Smoluchowski Institute of Physics, Jagiellonian
University,  PL-30348 Krak\'ow, Poland}

\author{K.~Hebeler}
\affiliation{Institut f\"ur Kernphysik, Technische Universit\"at 
Darmstadt, 64289 Darmstadt, Germany}
\affiliation{ExtreMe Matter Institute EMMI, GSI Helmholtzzentrum für Schwerionenforschung GmbH, 
64291 Darmstadt, Germany}

\author{T.~H\"uther}
\affiliation{Institut f\"ur Kernphysik, Technische Universit\"at 
Darmstadt, 64289 Darmstadt, Germany}

\author{H.~Kamada}
\affiliation{Department of Physics, Faculty of Engineering,
Kyushu Institute of Technology, Kitakyushu 804-8550, Japan}

\author{H.~Krebs}
\affiliation{Ruhr-Universit\"at
  Bochum, Fakult\"at f\"ur Physik und Astronomie, Institut f\"ur Theoretische Physik II, D-44780 Bochum, Germany}

\author{Ulf-G.~Mei{\ss}ner}
\affiliation{Helmholtz-Institut~f\"{u}r~Strahlen-~und~Kernphysik~and~Bethe~Center~for~Theoretical~Physics,
~Universit\"{a}t~Bonn,~D-53115~Bonn,~Germany}
\affiliation{Institut f\"ur Kernphysik, Institute for Advanced Simulation 
and J\"ulich Center for Hadron Physics, Forschungszentrum J\"ulich, 
D-52425 J\"ulich, Germany}
\affiliation{Tbilisi State University, 0186 Tbilisi, Georgia}

\author{J.A.~Melendez}
\affiliation{Department of Physics, The Ohio State University, 
Columbus, Ohio 43210, USA}

\author{A.~Nogga}
\affiliation{Institut f\"ur Kernphysik, Institute for Advanced Simulation 
and J\"ulich Center for Hadron Physics, Forschungszentrum J\"ulich, 
D-52425 J\"ulich, Germany}

\author{P.~Reinert}
\affiliation{Ruhr-Universit\"at
  Bochum, Fakult\"at f\"ur Physik und Astronomie, Institut f\"ur Theoretische Physik II, D-44780 Bochum, Germany}

\author{R.~Roth}
\affiliation{Institut f\"ur Kernphysik, Technische Universit\"at 
Darmstadt, 64289 Darmstadt, Germany}

\author{R.~Skibi\'nski}
\affiliation{M.~Smoluchowski Institute of Physics, Jagiellonian
University,  PL-30348 Krak\'ow, Poland}

\author{V.~Soloviov}
\affiliation{M.~Smoluchowski Institute of Physics, Jagiellonian
University,  PL-30348 Krak\'ow, Poland}

\author{K.~Topolnicki}
\affiliation{M.~Smoluchowski Institute of Physics, Jagiellonian
University,  PL-30348 Krak\'ow, Poland}

\author{J.P.~Vary}
\affiliation{Department of Physics and Astronomy, Iowa State
  University, Ames, Iowa 50011, USA}

\author{Yu.~Volkotrub}
\affiliation{M.~Smoluchowski Institute of Physics, Jagiellonian
University,  PL-30348 Krak\'ow, Poland}

\author{H.~Wita{\l}a}
\affiliation{M.~Smoluchowski Institute of Physics, Jagiellonian
University,  PL-30348 Krak\'ow, Poland}

\author{T.~Wolfgruber}
\affiliation{Institut f\"ur Kernphysik, Technische Universit\"at 
Darmstadt, 64289 Darmstadt, Germany}

\collaboration{LENPIC Collaboration}

%\collaboration{\color{red} EE: The list of authors is just a suggestion based %on the
%  (updated) list of authors of the previous LENPIC paper. Please check
%(also accommodations!) and adjust if necessary...}

\date{\today}

\begin{abstract}
We present a systematic investigation of few-nucleon systems and light nuclei using the
current LENPIC interactions comprising  semilocal momentum-space
regularized two- and three-nucleon forces up to third chiral order (N$^2$LO). Following our
earlier study utilizing the coordinate-space regularized interactions,
the two low-energy constants entering the three-body force are determined 
from the triton binding energy and the differential cross section
minimum in elastic nucleon-deuteron scattering. Predictions are made
for selected observables in elastic nucleon-deuteron scattering and in
the deuteron breakup reactions, for properties of the $A=3$ and $A=4$ nuclei, and for spectra of $p$-shell nuclei up to $A = 16$.  A comprehensive
error analysis is performed including an estimation of correlated truncation
uncertainties for nuclear spectra. The obtained predictions are
generally found to agree with experimental data within errors. Similar to the  
coordinate-space regularized chiral interactions at the same order,
a systematic overbinding of heavier nuclei is observed, which sets in
for $A \sim 10$ and increases with $A$. 
\end{abstract}

%\pacs{}

%\keywords{}

\maketitle

%%%%%%%%%%%%%%%%%%%%%%%%%%%%%%%%%%%%%%%%%%%%%%%%%%%%%%%%%%%%%%%%%%%%%%%%%%%%%
\section{Introduction} \label{Sec:intro}
A reliable quantitative first-principles description of nuclear
structure and reactions with quantified uncertainties remains one of the
main challenges in computational nuclear physics. Presently, the most
promising approach to reach this ambitious goal comprises a
combination of chiral effective field theory (EFT) to describe nuclear
interactions in harmony with the symmetries (and their breaking
pattern) of QCD with {\it ab initio} few-body methods to tackle the
quantum mechanical $A$-body problem. Remarkable progress has been
achieved in recent years in both lines of research, see 
Refs.~\cite{Epelbaum:2008ga,Machleidt:2011zz,Epelbaum:2019kcf,Hammer:2019poc,Tichai:2020dna,Marcucci:2019hml,Tews:2020hgp,Piarulli:2020mop,Lazauskas:2020jlw,Lee:2020meg,Gandolfi:2020pbj,Soma:2020xhv,Hergert:2020bxy}
for a selection of review articles on
these topics. At least for not-too-heavy nuclei, the accuracy of
theoretical predictions is, in most cases, limited by the uncertainties
of the nuclear interactions.

To address this challenge, the Low Energy
Nuclear Physics International Collaboration (LENPIC) aims at
developing accurate and precise two- and three-nucleon forces (3NF) by
pushing the EFT expansion to high chiral orders
and using these interactions to solve the
structure and reactions of light nuclei. In \cite{Binder:2015mbz,Maris:2016wrd,Binder:2018pgl}, we have already
explored selected nucleon-deuteron (Nd) scattering observables and the
structure of light- and medium-mass nuclei up to $A=48$ using the new generation of
the chiral EFT nucleon-nucleon potentials from
Refs.~\cite{Epelbaum:2014efa,Epelbaum:2014sza} up through fifth chiral 
order (N$^4$LO). The essential new feature of these interactions as compared to
the older potentials of Refs.~\cite{Epelbaum:2004fk,Entem:2003ft} and the new potentials developed by Entem
et al.~\cite{Entem:2017gor} is the usage of a \emph{local} regulator for
pion-exchange contributions, which allowed us to substantially reduce  finite-cutoff artifacts. These novel interactions have also 
been successfully tested in selected electroweak reactions with
two and three nucleons \cite{Skibinski:2016dve,Skibinski:2017vqs}.
While these exploratory studies employed NN interactions only and thus
should be regarded as incomplete starting from next-to-next-to-leading
order (N$^2$LO), the chiral order at which the 3NF starts to contribute,
they brought important new insights into the convergence pattern of
the chiral expansion. In particular, the resulting discrepancies between
theory and
experimental data were found to be in agreement with the expected size of
the missing 3NF contributions according to the Weinberg power
counting~\cite{Binder:2015mbz}.  

The expressions for the 3NF have been worked out completely up to fourth chiral
order (N$^3$LO) using dimensional regularization to deal with divergent
loop integrals~\cite{Ishikawa:2007zz,Bernard:2007sp,Bernard:2011zr};
see
also \cite{Krebs:2012yv,Girlanda:2011fh} for selected results at
N$^4$LO. A numerical implementation of the 3NF in the Faddeev and
Yakubovsky equations requires its partial-wave decomposition, which
can, in principle, be carried out in a brute-force way by numerically
performing the relevant angular integrations
\cite{Golak:2009ri,Hebeler:2015wxa}. However,  a coordinate-space
regulator for the long-range components of the 3NF, in line with the NN
potentials of Refs.~\cite{Epelbaum:2014efa,Epelbaum:2014sza}, was found to lead to numerical
instabilities when performing its partial-wave decomposition. While
this issue has been finally solved for the 3NF at tree level
(i.e.~at N$^2$LO) \cite{Epelbaum:2018ogq}, an extension of these studies to higher chiral
orders is a nontrivial task that would require further substantial efforts.  

These findings motivated the development of the semilocal
momentum-space regularized (SMS) NN potentials in
Ref.~\cite{Reinert:2017usi}, where both the short-range and long-range
contributions to the interaction are regularized in momentum
space. The other important differences to the semilocal
coordinate-space regularized (SCS) potentials of
Refs.~\cite{Epelbaum:2014efa,Epelbaum:2014sza} comprise the removal of
three redundant short-range operators at N$^3$LO and the usage of the
most up-to-date values of the pion-nucleon low-energy constants (LECs)
from the Roy-Steiner equation analysis of Ref.~\cite{Hoferichter:2015tha,Hoferichter:2015hva}. Moreover,
contrary to our earlier studies
\cite{Epelbaum:2014efa,Epelbaum:2014sza}
that relied on the Nijmegen partial-wave
analysis \cite{Stoks:1993tb}, the LECs accompanying the contact interactions have been
determined directly from the mutually consistent neutron-proton and
proton-proton scattering data of the 2013 Granada database \cite{Perez:2013jpa}.  
At the highest considered order N$^4$LO$^+$, where the ``$+$''
signifies the inclusion of four sixth-order contact interactions in
F-waves in order to describe certain very precise proton-proton
scattering data\footnote{The same short-range operators are also
  included in the N$^4$LO version of the potentials of
  Ref.~\cite{Entem:2017gor}.}, the NN potentials of Ref.~\cite{Reinert:2017usi}
allow for an outstanding description of the NN scattering data from
the 2013 Granada database below pion-production threshold.
In Ref.~\cite{Reinert:2020mcu}, these interactions have been extended by taking into account
isospin-breaking contributions up to N$^4$LO.  These are currently the most
precise chiral NN
interactions on the market, which for the intermediate cutoff value of
$\Lambda = 450$~MeV even qualify to be regarded as a partial-wave
analysis up to $E_{\hbox{\scriptsize lab}}=300$~MeV. These novel chiral EFT NN potentials have already been
successfully applied to Nd scattering~\cite{Epelbaum:2019zqc,Volkotrub:2020lsr} and to the $^2$H and $^3$He electroweak disintegration processes~\cite{Urbanevych:2020sjs}. 
They were also used in the recent
high-accuracy calculation of the electromagnetic form factors of the deuteron
\cite{Filin:2019eoe,Filin:2020tcs} and allowed, in particular, the prediction of
 the structure radius (quadrupole moment) of the deuteron with 
remarkable accuracy at the permille (percent) level. 

In this paper we present, for the first time,  results for
$p$-shell nuclei based on the SMS NN potentials of
Ref.~\cite{Reinert:2017usi} and also include the dominant 3NF at
N$^2$LO using the same regulator as employed in the two-body
interactions. We  employ the same convention
for the long-range 3NF as used in the NN interactions
by subtracting the locally regularized short-range terms to ensure that
the corresponding regularized three-nucleon potentials vanish at the
origin. Last but not least, we address the important issue of
estimating truncation errors for strongly correlated
observables such as the excitation energy spectra, see Ref.~\cite{Binder:2018pgl}
for a discussion.  

Our paper is organized as follows. In section~\ref{Sec:DetcDcE} we
provide the expressions for the regularized 3NF and
describe the determination of the LECs $c_D$ and $c_E$
from selected experimental data in the three-nucleon system.  Our
predictions for selected Nd scattering and breakup observables up to
N$^2$LO are summarized in section~\ref{sec:Nd_scattering}. Next,
sections~\ref{Sec:few-body_nuclei} and \ref{Sec:nuclei} are focused on
the properties of light nuclei with $A=3,4$ and on the energy spectra
of $p$-shell nuclei up to $A=16$, respectively. In section~\ref{Sec:correlated_spectra} we
perform an uncertainty analysis of the obtained predictions for the energy
spectra of light nuclei using a correlated error Bayesian model. This
allows us, for the first time, to reliably estimate the truncation
errors of our predictions for the excitation energies.   Finally, the most
important results of this study are summarized in section~\ref{conclusion}.

%%%%%%%%%%%%%%%%%%%%%%%%%%%%%%%%%%%%%%%%%%%%%%%%%%%%%%%%%%%%%%%%%%%%%%%%%%%%%
\section{Determination of \boldmath{$c_D$} and \boldmath{$c_E$}}
\label{Sec:DetcDcE}
Throughout this work, we employ the N$^2$LO three-nucleon force (3NF) regularized in momentum space in the same way
as the SMS two-nucleon interaction of Ref.~\cite{Reinert:2017usi}, namely
\beqa
\label{leading}
V^{\rm 3N}_\Lambda &=& \frac{g_A^2}{8 F_\pi^4}\;  e^{- \frac{\vec q_1^2 +
M_\pi^2}{\Lambda^2}}\, e^{- \frac{\vec q_3^2 + M_\pi^2}{\Lambda^2}}
\bigg\{
\frac{\vec \sigma_1 \cdot \vec q_1  \; \vec \sigma_3 \cdot \vec q_3}{(\vec  q_1^{\, 2} + M_\pi^2) \, (\vec q_3^{\, 2} + M_\pi^2)}
\nn
%&\times &\Big[T_{13}  \big( - 4 c_1 M_\pi^2 
%+ 2 c_3 \, \vec q_1 \cdot \vec
%    q_3 \big)  
&\times &\Big[T_{13}  \big( 2 c_3 \, \vec q_1 \cdot \vec q_3 - 4 c_1 M_\pi^2 \big)  
+  c_4 T_{132} \, \vec q_1 \times \vec q_3 
  \cdot \vec \sigma_2  \Big] \nn
%  &&{} \mbox{\hskip 3.75 true cm}+   
&+&
C\, \frac{\vec \sigma_1 \cdot \vec q_1}{\vec  q_1^{\, 2} + M_\pi^2}
\Big( 2 c_3 \, T_{13}\, \vec \sigma_3 \cdot \vec q_1 + c_4 T_{132} \; \vec q_1 \times \vec \sigma_3 
  \cdot \vec \sigma_2  \Big)   
  \nn
%  &&{} \mbox{\hskip 3.75 true cm}+   
&+&
C  \, \frac{\vec \sigma_3 \cdot \vec q_3}{\vec  q_3^{\, 2} + M_\pi^2}
\Big( 2 c_3 \, T_{13}\, \vec \sigma_1 \cdot \vec q_3 + c_4 T_{132}\; \vec \sigma_1 \times \vec q_3 
  \cdot \vec \sigma_2  \Big)  
  \nn
%  &&{}\mbox{\hskip 3.75 true cm}+
&+&
C^2 \, \Big( 2 c_3 \, T_{13} \, \vec \sigma_1 \cdot \vec \sigma_3 + c_4 T_{132} \; \vec \sigma_1 \times \vec \sigma_3 
  \cdot \vec \sigma_2 \Big)  \bigg\}\nn
&-& \frac{g_A \, D}{8 F_\pi^2}\; T_{13} \;
 e^{- \frac{\vec p_{12}^2 +\vec p_{12}^{\, \prime 2}
}{\Lambda^2}}\,
  e^{- \frac{\vec q_3^2 +
M_\pi^2}{\Lambda^2}}\, \bigg[
\frac{\vec \sigma_3 \cdot \vec q_3 }{\vec q_3^{\, 2} + M_\pi^2} \; 
\vec \sigma_1 \cdot \vec q_3  \nn
&+& 
C  \, \vec \sigma_1 \cdot \vec \sigma_3 \bigg] +  
\frac{1}{2} E \; T_{12} \, e^{- \frac{\vec p_{12}^2 +\vec p_{12}^{\, \prime 2}
  }{\Lambda^2}}\, e^{- \frac{3\vec k_{3}^2 +3\vec k_{3}^{\, \prime 2}
}{4\Lambda^2}} \nn
& +&  
\mbox{5 permutations}\,, 
\eeqa
where $\vec q_{i} = \vec p_i \, ' - \vec p_i$ is the momentum transfer of  nucleon $i$ with 
$\vec p_i \, '$ and $\vec p_i$ referring to the corresponding final and initial momenta, respectively,
$T_{ij} \equiv \fet \tau_i \cdot \fet \tau_j$, $T_{ijk} \equiv \fet
\tau_i \times \fet \tau_j \cdot \fet \tau_k$,
and $\vec \sigma_i$  ($\fet \tau_i$) are the Pauli
spin (isospin) matrices. We have also introduced the Jacobi momenta
$\vec p_{12} = (\vec p_1 - \vec p_2)/2$ and $\vec k_3 = 2 (\vec p_3 - (\vec p_1 + \vec p_2 )/2)/3$   in the initial state and 
$\vec p_{12}^{\, \prime} = (\vec p_1^{\, \prime} - \vec p_2^{\, \prime})/2$ and $\vec k_3^{\, \prime} = 2 (\vec p_3^{\, \prime} - (\vec p_1^{\, \prime} + \vec p_2^{\, \prime} )/2)/3$
in the final state. Further, $g_A$, $F_\pi$ and $M_\pi$ refer to the nucleon axial vector coupling, pion decay constant and pion mass, respectively, while the subtraction constant $C$ is given by \cite{Reinert:2017usi}
\beq
C  = -\frac{\Lambda \left(\Lambda ^2-2 M_\pi^2 \right) + 2 \sqrt{\pi } M_\pi^3 e^{\frac{M_\pi^2}{\Lambda ^2}}
   \text{erfc}\left(\frac{M_\pi}{\Lambda }\right)}{3 \Lambda ^3} \,,
 \eeq
where $\text{erfc} (x)$ is the complementary error function 
\beq
\text{erfc} (x) = \frac{2}{\sqrt{\pi}} \int_x^\infty dt \, e^{-t^2}\,.
\eeq
Finally, for the LECs $D$ and $E$, we employ the standard parametrization in terms of dimensionless
constants $c_D$ and $c_E$ via $D=c_D/(F_\pi^2 \Lambda_\chi)$ and
$E=c_E/(F_\pi^4 \Lambda_\chi)$ with $\Lambda_\chi = 700$~MeV. 

The subtraction terms proportional to $C$ in Eq.~(\ref{leading}) correspond to the {\it convention} employed in  Ref.~\cite{Reinert:2017usi}. It ensures, for example, that the locally regularized two-pion exchange 3NF in the curly brackets, Fourier transformed to coordinate space, vanishes at the origin (i.e.~for $\vec r_1 - \vec r_2 \to 0$ or  $\vec r_3 - \vec r_2 \to 0$) in order to minimize the admixture of short-range components. 
The applied regularization scheme, therefore, utilizes a local (nonlocal) regulator for long-range (short-range)
components of the 3NF. 

Partial-wave decomposition of the 3NF is accomplished numerically in momentum space in the usual way as described in detail in Refs.~\cite{Golak:2009ri, Hebeler:2015wxa, Hebeler:2020ocj}. Moreover, we have benchmarked the momentum-space results by independently carrying out the partial-wave decomposition in coordinate space. This way we have also explicitly verified that the subtracted long-range potentials vanish at the origin as required by our convention.  

We are now in the position to specify the values of the various LECs. For the pion-nucleon constants $c_i$,
we employ the values from matching chiral perturbation theory at next-to-leading order (NLO) in the NN-counting scheme
to the solutions of the Roy-Steiner equations for pion-nucleon scattering \cite{Hoferichter:2015tha, Hoferichter:2015hva}:
\beqa
c_1 &=& -0.74\mbox{ GeV}^{-1} \, , \nn
c_3 &=& -3.61\mbox{ GeV}^{-1} \, , \nn
c_4 &=& 2.44\mbox{ GeV}^{-1}  \, .
\eeqa
The same values are used in the SMS NN potentials of Ref.~\cite{Reinert:2017usi} at next-to-next-to-leading order (N$^2$LO). 

\begin{figure}[tb]
  \includegraphics[width=\columnwidth]{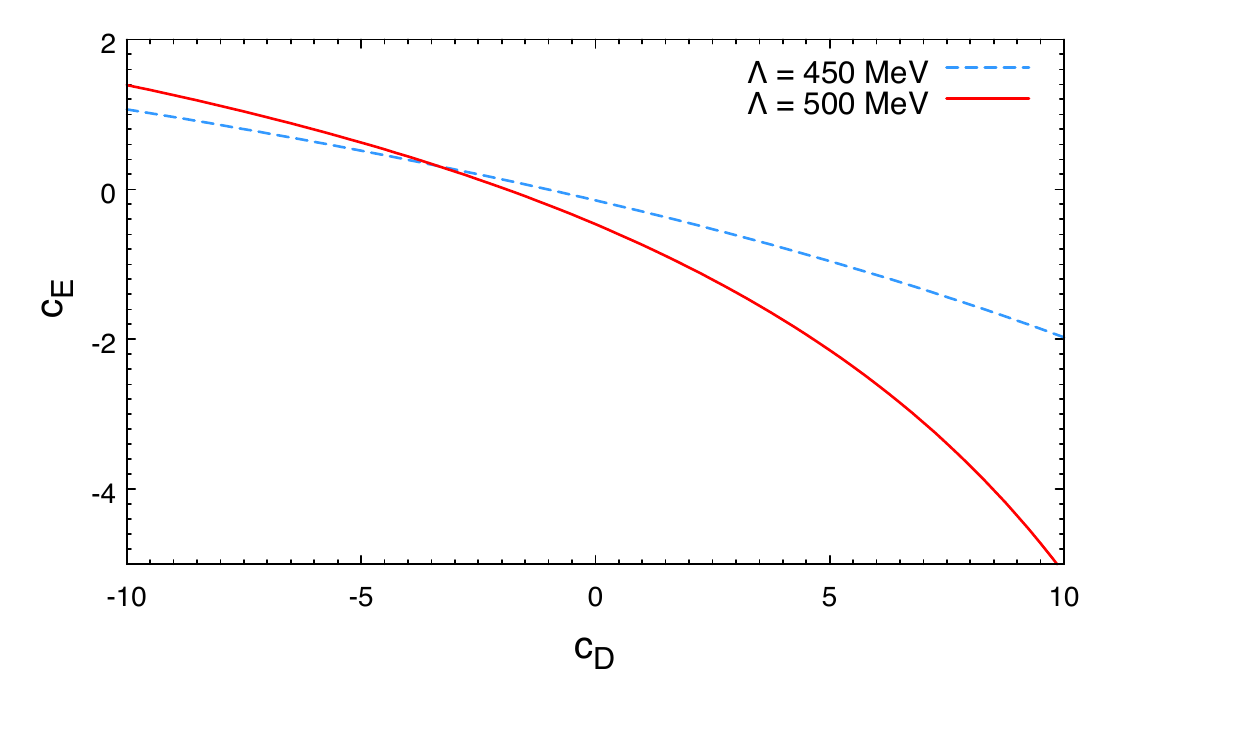}
%\vskip -0.7 true cm  
  \caption{\label{fig:cDcE} (Color online)
    Correlation between the LECs $c_D$ and $c_E$ induced by the
    requirement that the $^3$H binding energy be reproduced for the cutoff
    choices of $\Lambda = 450$~MeV (blue dashed line) and $\Lambda =500$~MeV (red solid
    line).}
\end{figure}

To determine the values of the LECs $c_D$, $c_E$, we require, following our previous studies \cite{Epelbaum:2002vt,Golak:2014ksa,Witala:2016inj,Epelbaum:2018ogq},
that the $^3$H binding energy is reproduced exactly. This constraint yields $c_E$ as a function of the LEC $c_D$ for
every value of the cutoff $\Lambda$. In Fig.~\ref{fig:cDcE}, we show the resulting $c_D$--$c_E$ correlations for the cutoff values
$\Lambda = 450$~MeV and $500$~MeV. As one may expect, the behavior is qualitatively similar to the
one found using the SCS interactions in Ref.~\cite{Epelbaum:2018ogq}. In particular, the larger momentum-space cutoff leads to a larger-in-magnitude negative slope of the function $c_E (c_D)$, exhibiting more
nonlinear behavior. 

Motivated by our findings in Ref.~\cite{Epelbaum:2018ogq}, the determination of the remaining LEC $c_D$ is carried out
by fitting the experimental data of Ref.~\cite{Sekiguchi:2002sf} for the differential cross section minimum at the nucleon beam energy
of $E_{N} = 70$~MeV. Specifically, the values of $c_D$ are determined from a least-squares fit of $12$ cross section data
points in the angular range of  $\theta_{\rm CM} \in [107.0^\circ, \, 140.4^\circ ]$ with the Coulomb force contribution
subtracted \cite{Delt}, and the statistical and systematic errors added in quadrature. This leads to the following central
values:
\beqa
\label{LECs}
&& c_D = \phantom - 2.485, \quad c_E = -0.528 \quad \mbox{for} \; \; \Lambda = 450\mbox{ MeV}, \nn
&& c_D = -1.626, \quad c_E = -0.063 \quad \mbox{for} \; \; \Lambda = 500\mbox{ MeV}.
\eeqa
The determination of uncertainties in the values of $c_D$, $c_E$ and their propagation will be considered in a separate publication.

\begin{figure*}[!tb]
  \includegraphics[width=\textwidth]{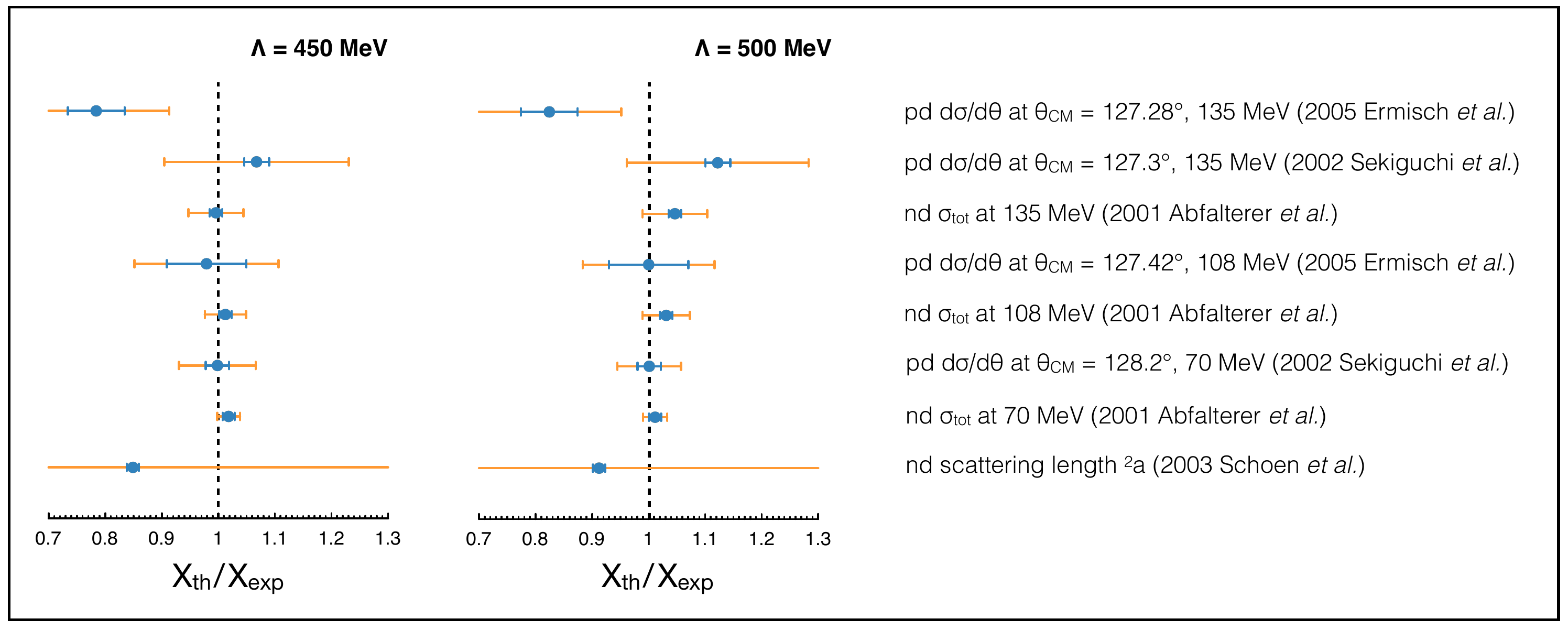}
%\vskip -0.7 true cm  
  \caption{\label{fig:cD_consistency} (Color online)
    Predictions for various Nd scattering observables based on the
    $c_D$ and $c_E$ values given in Eq.~(\ref{LECs}). For every
    considered observable $X$, a solid dot corresponds to the ratio of its
    calculated value $X_{\rm th}$ at N$^2$LO to the corresponding
    experimental value $X_{\rm exp}$. The smaller (blue) error bars
    correspond to the experimental relative uncertainty $\delta X_{\rm
    exp}/X_{\rm exp}$, where $\delta X_{\rm
    exp}$ includes both the statistical and systematic errors added in
  quadrature.  The larger (orange) error bars also take into account
  the estimated truncation error $\delta X_{\rm th}$ and correspond to
  $\sqrt{\delta X_{\rm
    exp}^2 + \delta X_{\rm
    th}^2}/X_{\rm exp}$. 
Experimental data are from \cite{Schoen:2003my} (2003 Schoen {\it et al.}),
\cite{Abfalterer:2001gw} (2001 Abfalterer {\it et al.}),
\cite{Sekiguchi:2002sf} (2002 Sekiguchi {\it et al.}) and \cite{Ermisch:2005kf} (2005 Ermisch {\it et al.}). 
}
\end{figure*}

%%%%%%%%%%%%%%%%%%%%%%%%%%%%%%%%%%%%%%%%%%%%%%%%%%%%%%%%%%%%%%%%%%%%%%%%%%%%%
\section{Nucleon-deuteron scattering} \label{sec:Nd_scattering}
We are now in the position to show selected results for
Nd scattering observables.
For a description of our formalism for solving the Faddeev-type
integral equations, see Ref.~\cite{Gloeckle:1995jg}. 
To estimate the truncation errors at N$^2$LO, we employ the Bayesian model $\bar C_{0.5-10}^{650}$ introduced
in Ref.~\cite{Epelbaum:2019zqc} based on the ideas of Refs.~\cite{Epelbaum:2014efa,Furnstahl:2015rha,Melendez:2017phj,Epelbaum:2019wvf}.  Specifically, for a three-nucleon
scattering observable $X ( E_N )$, we consider the chiral effective field theory (EFT) expansion up to N$^2$LO,
\beqa \label{eq:X_expansion}
X &=& X^{(0)} + \Delta X^{(2)} + \Delta X^{(3)} + \dots \nn
  &=:& X_{\rm ref} \left( c_0 + c_2 Q^2 + c_3 Q^3 + \dots \right) \,,
\eeqa
where $\Delta X^{(2)} := X^{(2)} - X^{(0)}$ and $\Delta X^{(3)} := X^{(3)} - X^{(2)}$, $Q$ is the expansion parameter,
the superscripts denote the chiral order $Q^n$, the ellipses refer to terms beyond N$^2$LO,
the quantity $X_{\rm ref} $ sets the overall scale and  
$c_i$ are the corresponding dimensionless coefficients. The reference scale $X_{\rm ref} $ is chosen using the
information from all three available chiral orders as described in Ref.~\cite{Epelbaum:2019zqc}. Assuming that
all dimensionless coefficients $c_i$ are normally distributed with the Gaussian prior  
\begin{equation} \label{eq:ci_prior}
  %\label{prior}
 {\rm pr} (c_i | \cbar ) = \frac{1}{\sqrt{2 \pi} \cbar } \, e^{-
   c_i^2/(2 \cbar ^2 )}\,,
\end{equation}  
and performing marginalization over the first $h=10$ neglected orders
for a uniform distribution of the hyperparameter $\cbar$ in the range of $\cbar \in [ 0.5, \, 10 ]$,
one obtains an analytical expression for the posterior probability distribution ${\rm pr}_h (\Delta | \{ c_i \} )$
for the dimensionless residual $\Delta_3 := c_4 Q^4 + \ldots + c_{3+h} Q^{3+h}$ to take a value $\Delta$
given the known coefficients $c_0$, $c_2$ and $c_3$, see Refs.~\cite{Epelbaum:2019zqc,Melendez:2017phj} for details.
This expression can be used to obtain truncation errors corresponding to any given degree-of-belief (DoB) interval.
Following Ref.~\cite{Epelbaum:2019zqc}, the expansion parameter $Q$ is defined according to  
\begin{equation}
  \label{eq:ExpParam}
Q = \max \left( \frac{p}{\Lambda_b}, \, \frac{M_\pi^{\rm
      eff}}{\Lambda_b} \right)\,,
\end{equation}
with $M_\pi^{\rm  eff} = 200$~MeV and the breakdown scale $\Lambda_b = 650$~MeV \cite{Epelbaum:2019wvf}. The momentum
$p$ is related to the laboratory energy $E_N$ of a given 3N scattering observable via \cite{Epelbaum:2019zqc}
\beq
\label{Scalep}
p = \sqrt{\frac{2}{3} m E_N}\,
\eeq
with $m$ denoting nucleon mass.

In our earlier paper describing results for the SCS
interactions \cite{Epelbaum:2018ogq}, the values of the LEC $c_D$ have been determined
from a combined fit to the Nd doublet scattering length $^2a$ as well as the total nd cross section
and the pd differential cross section data in the minimum region at the nucleon energies of $E_{N} = 70$~MeV, $108$~MeV and $135$~MeV.  
While we have used only the differential cross section data at the lowest energy of $70$~MeV in the present work, we show in
Fig.~\ref{fig:cD_consistency} that our $c_D$-determination is indeed consistent with the mentioned observables.

The only disagreement is observed for the differential cross section
data at $E_N = 135$~MeV measured at KVI \cite{Ermisch:2005kf}, which
also disagree with the data of Ref.~\cite{Sekiguchi:2002sf} at the same
energy, see the blue error bars in Fig.~\ref{fig:cD_consistency}.  
While our results indicate a slight preference for the experimental
data of  Ref.~\cite{Sekiguchi:2002sf}, the relatively large truncation
errors at N$^2$LO do not allow one to make a stronger conclusion at
this stage.

\begin{figure}[tb]
  \includegraphics[width=\columnwidth]{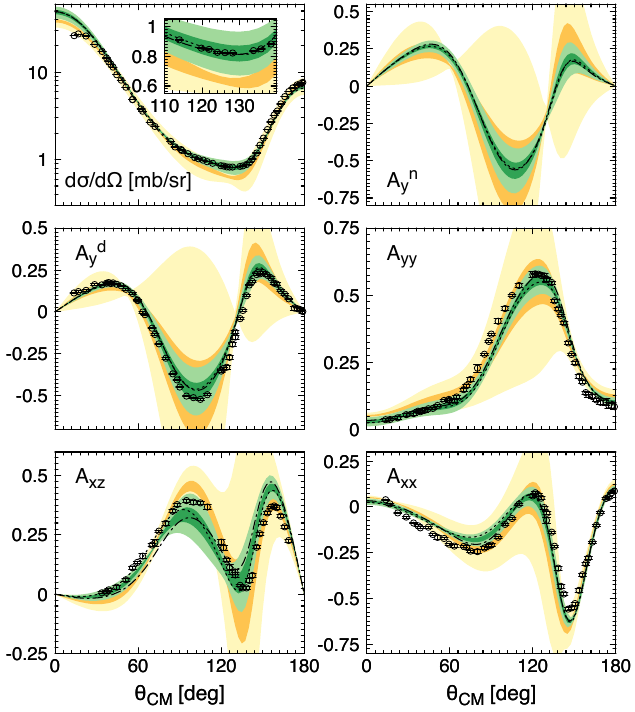}
%\vskip -0.7 true cm  
  \caption{\label{fig:Nd_elastic} (Color online)
 Results for the differential cross section, nucleon and
  deuteron analyzing powers $A_y^n$ and $A_y^d$ as well as deuteron
  tensor analyzing powers $A_{yy}$, $A_{xz}$ and $A_{xx}$ in elastic
  nucleon-deuteron scattering  at laboratory energy of
  $E_N = 70$~MeV  at NLO
  (yellow bands) and N$^2$LO (green bands) for $\Lambda =
  450$~MeV. Dotted lines show the N$^2$LO results based on
  the SMS NN forces from Ref.~\cite{Reinert:2017usi} accompanied with the {\it
    unsubtracted} (i.e.~with $C(M_\pi) = 0$) 3NF from Ref.~\cite{Epelbaum:2019zqc},
  while dashed-dotted lines are N$^2$LO predictions based on the
  SCS (NN+3NF) interactions from Ref.~\cite{Epelbaum:2018ogq}.  
The light (dark) shaded bands indicate $95\%$ ($68\%$)
DoB intervals using the Bayesian model $\bar C_{0.5-10}^{650}$. 
  Open circles are proton-deuteron data from
  Ref.~\cite{Sekiguchi:2002sf}. 
}
\end{figure}

In Fig.~\ref{fig:Nd_elastic}, we show our NLO and N$^2$LO predictions
for selected observables in elastic Nd scattering at $E_N = 70$~MeV. 
Notice that the truncation errors are symmetric, and the actual
results of our calculation lie in the middle of the corresponding
error bands. At both NLO and N$^2$LO, the experimental data are, in
most cases, reasonably well described by our calculations. We also
compare our results to those of Ref.~\cite{Epelbaum:2019zqc} based on the same NN
interactions but using the unsubtracted version of the 3NF, as well as
to our earlier results using the SCS two-~\cite{Epelbaum:2014efa} and
three-nucleon forces from Ref.~\cite{Epelbaum:2018ogq} at the regulator value $R=0.9$~fm.   It is
reassuring to see that all shown N$^2$LO results agree with each other
within errors.

\begin{figure}[tb]
  \includegraphics[width=0.99\columnwidth]{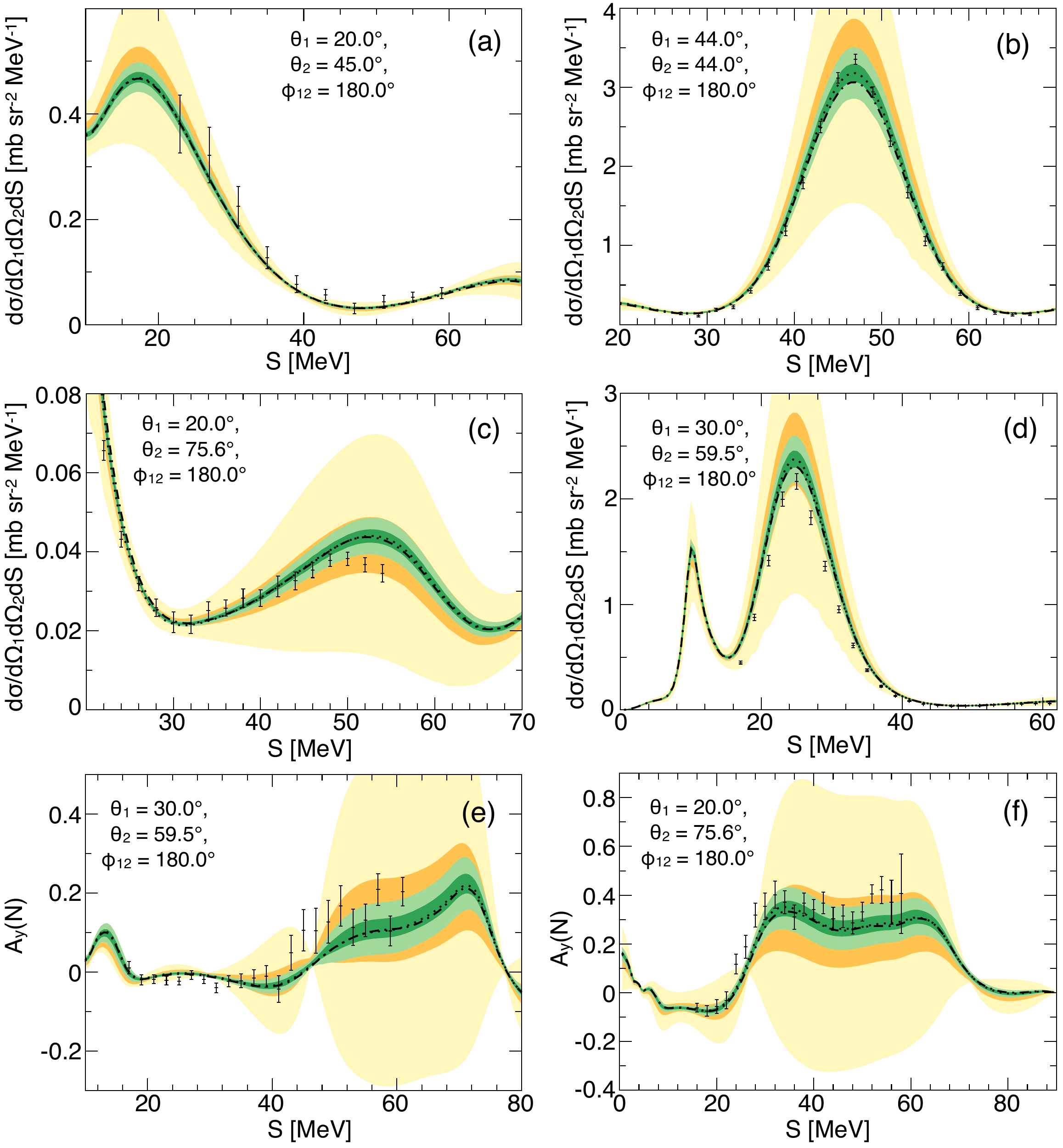}
%\vskip -0.7 true cm  
  \caption{\label{fig:Nd_breakup} (Color online)
 Results for the differential cross section, (a)-(d), and nucleon
  vector analyzing power $A_y$ as functions of the kinematical locus
  variable $S$ for the deuteron breakup reaction 
at laboratory energy of
  $E_N = 65$~MeV  at NLO
  (yellow bands) and N$^2$LO (green bands) for $\Lambda =
  450$~MeV. Proton-deuteron data for (a), (c), (f) are taken from Ref.~\cite{Bodek1}
while those shown in (b), (d) and (e) are from Ref.~\cite{Bodek2}.  For remaining notation see Fig.~\ref{fig:Nd_elastic}.}
\end{figure}

We have also calculated selected breakup observables at $E_N =
65$~MeV, for which experimental data are available. In
Fig.~\ref{fig:Nd_breakup}, we show the five-fold differential cross section and
nucleon vector analyzing power $A_y$ as functions of the kinematical
locus variable $S$ for selected configurations specified by the
detection angles $\theta_1$, $\theta_2$ and $\phi_{12}$ in the
laboratory system; see
Ref.~\cite{Gloeckle:1995jg} for the definition of kinematical
variables,  which may serve as representative examples.

One observes a similar picture as for the considered elastic
scattering observables. In particular, the experimental data are well
reproduced, and our N$^2$LO results agree well with those obtained
both using the SCS (NN+3NF) interactions and the SMS forces with
unsubtracted 3NF. Furthermore, our N$^2$LO results for the
differential cross sections in Fig.~\ref{fig:Nd_elastic} agree well
with the predictions based on the CD Bonn potential, see Fig.~6 of
Ref.~\cite{Skibinski:2006cg}. This should not come as a surprise since 3NF effects
appear to be fairly small for the considered cases. Notice,
however, that relativistic corrections turn out to
be non-negligible for the cross section in the panels (b) and
(d). In Ref.~\cite{Skibinski:2006cg}, they were found to decrease the predictions for
the differential
cross section around the maximum by almost $10\%$.  In chiral EFT,
relativistic corrections to the Nd scattering amplitude need to be taken into account starting
from N$^3$LO. Their expected size is, therefore, in qualitative
agreement with the estimated size of the neglected N$^3$LO
contributions as reflected by the width of the green uncertainty
bands.   Last but not least, we have also calculated breakup
configurations considered in Ref.~\cite{KurosZonierczuk:2002uw}, which
feature more pronounced 3NF effects. In all considered cases (not
shown here), we found similar results to the
ones based on high-precision phenomenological NN potentials in
combinations with the Urbana IX \cite{Pudliner:1997ck} and the updated
Tucson-Melbourne \cite{Coon:2001pv} 3NF
models. 

\begin{table*}[t]
  \begin{ruledtabular}
    \begin{tabular*}{\textwidth}{@{\extracolsep{\fill}}llcrrrrrrrrrrrrr}
%\begin{tabular}{llcrrrrrrrrrrrrr}
         &                                   & $\Lambda$ &  E &    $\langle H \rangle$ &   $\langle T \rangle$&  $\langle V_{NN} \rangle$ &  $\langle V_{3NF} \rangle$ &  $\langle T_{CSB} \rangle$ &  $\langle \Psi | \Psi \rangle$ & P(S) & P(P) &  P(D) & r(p) & r(n) \\
\hline
 \multirow{4}{*}{$^3$H}   &   LO    &  \multirow{4}{*}{450}   &   $-$12.22 &      $-$12.24 &  52.38 &   $-$64.61 &   ---      & $-$10.505 &  1.0000 &   96.25 &   0.019 &    3.73 &    1.250 &    1.319 \\
   &   NLO                &                                   &   $-$8.515 &      $-$8.521 &  34.31 &   $-$42.82 &   ---      &  $-$5.798 &  0.9999 &   94.79 &   0.028 &    5.19 &    1.556 &    1.702 \\
    &   N$^2$LO (NN-only)  &                                  &   $-$8.143 &      $-$8.148 &  34.94 &   $-$43.08 &   ---      &  $-$5.552 &  0.9998 &   93.28 &   0.044 &    6.68 &    1.595 &    1.752 \\
    &   N$^2$LO+3NFs       &                                  &   $-$8.483 &      $-$8.489 &  36.13 &   $-$44.16 &    $-$0.459 &  $-$5.840 &  0.9995 &   92.54 &   0.077 &    7.38 &    1.576 &    1.725 \\
\hline
\multirow{4}{*}{$^3$H}   &   LO  &  \multirow{4}{*}{500}      &   $-$12.52 &      $-$12.53 &  57.84 &   $-$70.36 &   ---      & $-$11.528 &  0.9999 &   94.96 &   0.036 &    5.01 &    1.224 &    1.286 \\
   &   NLO                &                                   &   $-$8.325 &      $-$8.332 &  35.87 &   $-$44.19 &   ---      &  $-$6.150 &  0.9998 &   94.29 &   0.032 &    5.68 &    1.575 &    1.725 \\
    &   N$^2$LO (NN-only)  &                                  &   $-$7.920 &      $-$7.926 &  37.94 &   $-$45.86 &   ---      &  $-$5.754 &  0.9996 &   92.06 &   0.059 &    7.89 &    1.625 &    1.787 \\
    &   N$^2$LO+3NFs       &                                  &   $-$8.482 &      $-$8.488 &  40.27 &   $-$48.09 &   $-$0.660  &   $-$6.24 &  0.9992 &   91.39 &   0.109 &    8.50 &    1.581 &    1.731 \\
\hline
 \multirow{4}{*}{$^3$He}   &      LO  &  \multirow{4}{*}{450} &   $-$11.34 &      $-$11.33 &  51.45 &   $-$62.79 &   ---      &   9.851 &  1.0000 &   96.24 &   0.019 &    3.75 &    1.342 &    1.264 \\
   &   NLO               &                                    &   $-$7.751 &      $-$7.745 &  33.55 &   $-$41.30 &   ---      &   5.217 &  0.9998 &   94.79 &   0.027 &    5.18 &    1.744 &    1.579 \\
    &   N$^2$LO (NN-only) &                                   &   $-$7.397 &      $-$7.392 &  34.15 &   $-$41.55 &   ---      &   4.967 &  0.9998 &   93.29 &   0.043 &    6.67 &    1.797 &    1.620 \\
   &   N$^2$LO+3NFs      &                                    &   $-$7.734 &      $-$7.729 &  35.37 &   $-$42.65 &   $-$0.452  &    5.26 &  0.9995 &   92.57 &   0.076 &    7.35 &    1.766 &    1.598 \\
\hline
 \multirow{4}{*}{$^3$He}   &   LO  &  \multirow{4}{*}{500}    &   $-$11.63 &      $-$11.62 &  56.88 &   $-$68.51 &   ---      &  10.865 &  0.9999 &   94.94 &   0.036 &    5.02 &    1.308 &    1.237 \\
   &   NLO               &                                    &   $-$7.574 &      $-$7.568 &  35.07 &   $-$42.65 &   ---      &   5.557 &  0.9997 &   94.30 &   0.031 &    5.67 &    1.768 &    1.598 \\
   &   N$^2$LO (NN-only) &                                    &   $-$7.194 &      $-$7.188 &  37.11 &   $-$44.30 &   ---      &   5.164 &  0.9995 &   92.08 &   0.059 &    7.86 &    1.834 &    1.649 \\
   &   N$^2$LO+3NFs      &                                    &   $-$7.739 &      $-$7.733 &  39.44 &   $-$46.54 &   $-$0.641  &    5.65 &  0.9991 &   91.43 &   0.107 &    8.47 &    1.772 &    1.602 
    \end{tabular*}
\caption{\label{tab:3H3He} Summary of energies and wave function properties for $^3$H/$^3$He. See text for explanations. Energies and cutoffs are given in MeV except for $\langle T_{CSB} \rangle$ which is given in keV. Radii are given in fm and the S,P and D-state probabilities are given in \%. }
\end{ruledtabular}   
\end{table*}

%%%%%%%%%%%%%%%%%%%%%%%%%%%%%%%%%%%%%%%%%%%%%%%%%%%%%%%%%%%%%%%%%%%%%%%%%%%%%
\section{A=3 and 4 nuclei} \label{Sec:few-body_nuclei}
With the interactions specified in the previous section, we now calculate the ground state
energies and excitation spectra up to the  $p$-shell.  In this section, we
focus on the $A=3$ and $A=4$ bound states, for which we solve Faddeev and Yakubovsky equations in momentum space
as outlined in Ref.~\cite{Binder:2018pgl}. The addition of 3NFs has
been discussed in Ref.~\cite{Nogga:2001cz}.

\begin{table*}[t]
    \begin{ruledtabular}
    \begin{tabular*}{\textwidth}{@{\extracolsep{\fill}}llcrrrrrrrrrrrrr}
%\begin{tabular}{llcrrrrrrrrrrrrr}
         &                                   & $\Lambda$ &  E &    $\langle H \rangle$ &   $\langle T \rangle$&  $\langle V_{NN} \rangle_1$ &  $\langle V_{NN} \rangle_2$ &  $\langle V_{3NF} \rangle$ &  $\langle \Psi | \Psi \rangle_1$ & $\langle \Psi | \Psi \rangle_2$ & P(S) & P(P) &  P(D) & r(p)/r(n) \\
\hline
 \multirow{4}{*}{$^4$He}   &   LO    &  \multirow{4}{*}{450} & $-$49.99 &    $-$49.98 &  124.4 &   $-$174.4 &   $-$174.4 &       --- &  0.99952 &  0.99980 &   95.71 &    0.070 &    4.21 &   0.990 \\
   &   NLO                &                                  & $-$29.36 &    $-$29.34 &  71.47 &   $-$100.8 &   $-$100.8 &       --- &  0.99914 &  0.99932 &   92.02 &    0.129 &    7.84 &   1.375 \\
    &   N$^2$LO, NN-only  &                                 & $-$27.32 &    $-$27.28 &  71.95 &    $-$99.3 &    $-$99.2 &       --- &  0.99887 &  0.99894 &   89.71 &    0.211 &   10.07 &   1.423 \\
    &   N$^2$LO+3NFs       &                                 & $-$28.62 &    $-$28.59 &  75.73 &   $-$102.0 &   $-$102.0 &    $-$2.376 &  0.99934 &  0.99923 &   86.72 &    0.462 &   12.81 &   1.423 \\
\hline							                                                                                                                                               
\multirow{4}{*}{$^4$He}   &   LO  &  \multirow{4}{*}{500}    & $-$51.47 &    $-$51.46 &  139.2 &   $-$190.7 &   $-$190.7 &       --- &  0.99943 &  0.99973 &   93.73 &    0.147 &    6.11 &   0.955 \\
   &   NLO                &                                  & $-$28.15 &    $-$28.12 &  74.56 &   $-$102.7 &   $-$102.7 &       --- &  0.99872 &  0.99863 &   90.96 &    0.154 &    8.87 &   1.408 \\
    &   N$^2$LO, NN-only  &                                 & $-$25.95 &    $-$25.85 &  78.54 &   $-$104.5 &   $-$104.4 &       --- &  0.99806 &  0.99758 &   87.36 &    0.303 &   12.32 &   1.469 \\
    &   N$^2$LO+3NFs       &                                 & $-$28.72
                                                              &
                                                                $-$28.62
                                                                                       &  86.71 &   $-$111.9 &   $-$111.9 &    $-$3.474 &  0.99873 &  0.99811 &   85.06 &    0.597 &   14.34 &   1.424
     \end{tabular*}                                                                                                                                                                                        
\caption{\label{tab:4He} Summary of energies and wave function
  properties for $^4$He. See text for explanations. Energies and cutoffs
  are given in MeV. The radius is given in fm and the S, P and D-state
  probabilities are given in \%. }
\end{ruledtabular}   
\end{table*}

For $A=3$, we reach a numerical accuracy of 1~keV for the binding energy and the expectation values.
For practical calculations, the number of partial-wave channels needs to be truncated. To reach
the desired accuracy, we take  into account all partial waves with two-body subsystem angular momentum
less than or equal to 5. This includes a small admixture of total isospin $T=3/2$ states
in $^3$H and $^3$He. It has been shown in \cite{Nogga:2002qp} that this is necessary to reach the
same accuracy for the expectation values of the Hamiltonian.

In $^3$He, additionally, the point Coulomb interaction has been included for the $pp$
subsystem. For the numerical calculation in momentum space,
the Coulomb force has been Fourier transformed using a cutoff at distances of
20~fm. It has been checked that larger distances do not contribute to the 3N
binding energies.

Our results are summarized in Table~\ref{tab:3H3He} for LO, NLO and N$^2$LO interactions.
For N$^2$LO, we also give results for NN interactions only.

It can been seen that the binding energy $|E|$ based purely on NN interactions
decreases for both cutoffs when the chiral order increases. At N$^2$LO,
an attractive contribution of the 3NF increases the binding energy
and brings it, by construction, in agreement with the experimental value of
$-$8.482~MeV for $^3$H~\cite{Wang_2017}. For the calculation of the energies, we used an averaged
proton and neutron mass of 938.918~MeV. The slight deviation of the expectation
values $\langle H \rangle$ from the binding energy is due to the additional contribution
$\langle T_{CSB} \rangle$ resulting from employing physical proton and neutron
masses. Due to the larger binding energy this effect is more pronounced at LO.
The approximately $\pm 6$~keV for $^3$He and $^3$H for NLO and N$^2$LO
are in line with the results for phenomenological interactions \cite{Nogga:2002qp}.
Including this contribution, we find in LO, NLO and N$^2$LO (including the 3NF) for
$\Lambda=500$~MeV  a charge
symmetry breaking (CSB) difference of the binding energies of 910, 764 and 755~keV comparable
to the experimental value of 764~keV. Results for $\Lambda=450$~MeV are very similar.

It is well known that the partial-wave convergence of Faddeev components
for the 3N system is faster than the convergence of the wave functions. Therefore, it is
advantageous to normalize the overlap of Faddeev components and wave functions \cite{Binder:2018pgl}.
Missing high partial waves then lead to a deviation of the norm $\langle \Psi | \Psi \rangle$
of the 3N system from one. The size of this effect is larger when
higher orders of the interactions or 3NFs are employed, indicating that these
interactions induce contributions to higher partial waves. In all cases, our
truncation of the partial-wave basis provides more than 99.9\% off the norm. 

Table~\ref{tab:3H3He} also summarizes the S-, P- and D-state probabilities. As usual, the
contribution of the P-state is small. The D-state visibly contributes, where again
the higher orders and especially 3NFs lead to an increasing D-state probability.
This is in line with results based on different interactions \cite{Nogga:2001cz}.

Finally, we also give values for point proton and neutron matter radii. These have been obtained
based on a Fourier transform of the wave functions. The observed
pattern showing increasing (decreasing) values of the radii with the chiral order
(when adding the 3NF) are qualitatively in line with the changes in
binding energy. Note however that 3NFs break this correlation to some extent.

For $A=4$, we solve the set of Yakubovsky equations. Since three orbital angular momenta 
contribute to the partial-wave expansion, we need to constrain at least two of 
them to end up with a finite number of partial waves. For the calculations done here, 
we again use the two-body subsystem angular momentum to less than or equal to 5. Additionally, 
we only use orbital angular momenta less than or equal to 6 and also constrain the sum of 
all orbital angular momenta to 10 or less. We also use isospin symmetry and therefore only take the by-far dominant isospin $T=0$ channels into account. We checked that these constraints lead to an uncertainty of approximately 10~keV for the binding energies and 50~keV for the expectation values. Since we only take isospin $T=0$ channels into account, there is no contribution from charge-symmetry breaking to the kinetic energy. Also the proton and neutron radii are exactly equal in this approximation. 

The pattern of binding is very similar to the $A=3$ nuclei, as can be seen in Table~\ref{tab:4He}. In leading order, there is strong overbinding compared to
the experimental $^4$He binding energy of $-28.296$~MeV~\cite{Wang_2017}.  
This is drastically reduced at NLO, leading even to a slight 
underbinding for $\Lambda=500$~MeV. At N$^2$LO, $^4$He is clearly underbound for both 
cutoffs when only NN interactions are used. Adding the 3NFs then leads to mild overbinding again. The effect of the 3NFs is larger for the larger cutoff. 
An alternative approach has been proposed~\cite{SanchezSanchez:2020kbx}
which reduces the strong overbinding at LO and it will be interesting to see if that approach
is successful when consistent higher orders are developed.

As described in more detail in \cite{Nogga:2001cz}, the $A=4$ calculations are solved 
using two kinds of Jacobi coordinates that either single out a three-nucleon subsystem 
(3+1) and a spectator nucleon or two two-nucleon subsystems (2+2). It is advantageous to use both kinds of coordinates simultaneously since this allows for the most effective representation of the Yakubovsky components. Again, like in $A=3$, we 
perform the normalization of the wave function using overlaps of the 
Yakubovsky component and the wave function and then calculate the norm using 
only the wave function in either the 3+1 representation ($\langle \Psi| \Psi \rangle_1 $)
or in the 2+2 one ($\langle \Psi| \Psi \rangle_2 $). As one can see in the table, the deviation from 1 is less than $1\permil$ for both representations of the
wave function.

We also give the expectation values of the Hamiltonian, the kinetic energy and 
the potential energy in both representations. In contrast to $A=3$, there are 
small deviations between the expectation value $\langle H \rangle$ and $E$ that 
can not be traced back to additional charge-symmetry breaking contributions in the 
kinetic energy but are due to the missing isospin $T=1$ and $T=2$ components 
and due to missing higher partial waves in intermediate states. 
For $A=3$, such deviations did not show up because all isospin channels 
were included and because no intermediate steps were necessary to 
compute wave functions and expectation values. 

Table~\ref{tab:4He} also compiles the S-, P- and D-state probabilities, which 
follow a very similar trend as the ones for $A=3$. Finally, the point proton/neutron matter
radius is also given. Again, we observe the expected correlation with the binding 
energy with the exception that 3NFs can contribute additional binding without 
decreasing the radius.

We conclude this section with a short discussion on the radii of the
$A=3,4$ nuclei. The (unobservable) point-proton matter radius $r_p$ quoted in Tables~\ref{tab:3H3He} and \ref{tab:4He} is related to the charge radius $r_c$ by the well known
equation \cite{Lu:2013ena}
\begin{equation}
r_c^2 = r_p^2 + R_p^2 + \frac{3}{4 m_p^2} + \frac{N}{Z} R_n^2 +
r_{\rm so}^2 + r_{\rm mec}^2 + \dots \,,
\end{equation}  
where $m_p$, $R_p$ and $R_n$ are the proton mass, charge radius and 
the neutron charge radius, respectively, while the ellipses refer to higher-order
relativistic corrections. Further, $r_{\rm so}^2$ denotes 
the contribution of the spin-orbit term of  relativistic nature,
see \cite{Friar:1997js,Filin:2020tcs} for more details, while $r_{\rm mec}^2$ refers to the
contribution of the exchange charge density. In a close analogy to
the deuteron, see e.g.~\cite{Jentschura:2011}, one can define the point-proton structure radius $r_{\rm
  str}$ via 
\begin{equation}
r_{\rm
  str}^2 = r_c^2 - \left( R_p^2 + \frac{3}{4 m_p^2} + \frac{N}{Z}
  R_n^2 \right) \,.
\end{equation}  
This quantity is observable and differs from the point-proton matter
radius $r_p$ by taking into account the contributions associated with nuclear
binding mechanisms such as $r_{\rm so}^2$ and $r_{\rm mec}^2$.   
Since these corrections first appear at N$^3$LO, see
Refs.~\cite{Kolling:2009iq,Kolling:2011mt,Krebs:2019aka}, our N$^2$LO
predictions for $r_p$ coincide with the ones for the structure radii, whose experimental values
can be extracted from the corresponding charge radii 
$r_{c, \rm \, 3H}^{\rm exp} = 1.755(86)$~fm \cite{Amroun:1994qj},
$r_{c, \rm \, 3He}^{\rm exp} = 1.973(14)$~fm \cite{Sick:2014yha} and 
$r_{c, \rm \, 4He}^{\rm exp} = 1.681(4)$~fm \cite{Sick:2008zza}. Using the
CODATA-2018 recommended value for the proton radius, $R_p = 0.8414(19)$~fm
\cite{CODATA2018}, along with the
current PDG value for the square charge radius of the neutron, $R_n^2
= -0.1161(22)$~fm$^2$ \cite{Zyla:2020zbs}\footnote{Notice, however, that our recent
  determination of $R_n^2$ from the atomic isotope shift data \cite{Filin:2019eoe,Filin:2020tcs} yielded
  a somewhat smaller in magnitude value of $R_n^2
= -0.105^{+0.005}_{-0.006}$~fm$^2$, which would result in slightly
smaller structure radii.}, we extract the corresponding structure
radii $r_{\rm str, \, 3H}^{\rm exp} = 1.604(96)$~fm, $r_{\rm str, \,
  3He}^{\rm exp} = 1.792(17)$~fm and $r_{\rm str, \,
  4He}^{\rm exp} = 1.484(6)$~fm. Notice that this value for $r_{\rm str, \,
  4He}^{\rm exp}$ is significantly (by $\sim 1.5\%$) larger than the
one of $r_{\rm str, \,  4He}^{\rm exp} = 1.462(6)$~fm given in
Ref.~\cite{Lu:2013ena} and also quoted in our earlier studies \cite{Binder:2015mbz,Epelbaum:2018ogq}. This
difference is caused entirely by employing the updated (smaller) value for the
proton radius recommended by the CODATA group \cite{CODATA2018}, and consistent with decade-long findings using dispersion theory, see e.g. \cite{Hammer:2019uab}.

Our N$^2$LO predictions underestimate the central experimental values for the
structure radii of all three considered nuclei. The amount of underestimation
is slightly smaller for the larger cutoff $\Lambda = 500$~MeV. This is
in line with the recent high accuracy calculations of the deuteron
charge and quadrupole form factors \cite{Filin:2019eoe,Filin:2020tcs}, where the contribution of
the two-body short-range charge density was found to decrease with
increasing cutoff values. Using the same Bayesian model $\bar
C_{0.5-10}^{650}$ as employed in section
\ref{sec:Nd_scattering} for 3N scattering observables to estimate the truncation errors for the radii, our N$^2$LO
predictions for $\Lambda= 500$~MeV 
$r_{\rm str, \, 3H}^{\rm N2LO} = 1.581(29)$~fm, 
$r_{\rm str, \, 3He}^{\rm N2LO} = 1.772(37)$~fm and
$r_{\rm str, \, 4He}^{\rm N2LO} = 1.424(36)$~fm
are found to be consistent with the experimental values for the $A=3$ nuclei.
At the $95\%$ Bayesian confidence level corresponding to $r_{\rm str,
  \, 4He}^{\rm N2LO} = 1.424(109)$~fm, our result for $^4$He is also
in agreement with the experimental datum.

%%%%%%%%%%%%%%%%%%%%%%%%%%%%%%%%%%%%%%%%%%%%%%%%%%%%%%%%%%%%%%%%%%%%%%%%%%%%%
\begin{table*}
   \begin{ruledtabular}
    \begin{tabular*}{\textwidth}{@{\extracolsep{\fill}}lcc|dddd|d}
%  \begin{tabular}{lcc|dddd|d}
&&&&&&&\\[-12pt]
      $^A$Z$(J^{\pi},T)$ & $\Lambda$ & $\alpha$ (fm$^{4}$) 
                       & \multicolumn{1}{c}{LO}
                       & \multicolumn{1}{c}{NLO}
                       & \multicolumn{1}{c}{N$^2$LO, NN-only}
                       & \multicolumn{1}{c|}{N$^2$LO NN+3NFs}
                       & \multicolumn{1}{c}{Exp. (MeV)} \\
    \hline
    \hline
    $^4$He$(0^+, 0)$   & 450 & 0.04  &  -49.891(2)&  -29.339(3) &  -27.254(5) &  -28.447(4) & -28.296 \\
                       & 450 & 0.08  &  -49.733(1)&  -29.366(1) &  -27.260(2) &  -28.527(2) & \\
                       & 500 & 0.04  &  -51.327(1)&  -28.087(3) &  -25.816(5) &  -28.585(5) & \\
                       & 500 & 0.08  &  -51.167(1)&  -28.123(2) &  -25.807(3) &  -28.630(2) & \\
    \hline
    $^6$He$(0^+, 1)$   & 450 & 0.04  &  -46.5(5)  &  -28.73(16) &  -27.09(16) &  -28.84(20) & -29.27 \\
                       & 450 & 0.08  &  -46.7(3)  &  -27.86(14) &  -27.18(10) &  -29.04(7)  & \\
                       & 500 & 0.04  &  -47.2(6)  &  -27.27(15) &  -25.66(16) &  -29.08(20) & \\
                       & 500 & 0.08  &  -47.6(4)  &  -27.39(10) &  -25.69(7)  &  -29.21(6)  & \\
    $E_x(2^+, 0)$      & 450 & 0.08  &    3.5(9)  &    2.10(31) &    2.09(23) &    2.10(15) &   1.80 \\
                       & 500 & 0.08  &    3.6(1.0)&    2.08(23) &    2.08(12) &    2.08(07) & \\
    \hline
    $^6$Li$(1^+, 0)$   & 450 & 0.04  &  -50.1(4)  &  -31.79(11) &  -30.20(16) &  -31.85(15) & -31.99 \\
                       & 450 & 0.08  &  -50.4(3)  &  -31.93(9)  &  -30.28(6)  &  -32.04(6)  & \\
                       & 500 & 0.04  &  -50.7(6)  &  -30.33(12) &  -28.75(15) &  -32.17(20) & \\
                       & 500 & 0.08  &  -51.1(3)  &  -30.45(6)  &  -28.77(5)  &  -32.29(4)  & \\
    $E_x(2^+, 0)$      & 450 & 0.08  &    5.3(8)  &    2.90(17) &    2.85(12) &    2.40(7)  &   2.19 \\ 
                       & 500 & 0.08  &    5.1(9)  &    2.93(14) &
                                                                  2.82(7)
                       &    2.41(7)  & \\[-2pt]
 %     \hline
    %
  \end{tabular*}
  \caption{\label{Tab:res_A4A6}
    ground state energies of $^4$He, $^6$He, and $^6$Li from LO up
    through N$^2$LO including 3NFs $\alpha=0.08$ and $0.04$~fm$^4$
    Quoted uncertainties are the extrapolation uncertainties only. Energies and cutoffs
  are given in MeV.  Experimental values are from Refs.~\cite{Wang_2017, TILLEY20023}.}
\end{ruledtabular}   
\end{table*}
%%%%%%%%%%%%%%%%%%%%%%%%%%%%%%%%%%%%%%%%%%%%%%%%%%%%%%%%%%%%%%%%%%%%%%%%%%%%%

%%%%%%%%%%%%%%%%%%%%%%%%%%%%%%%%%%%%%%%%%%%%%%%%%%%%%%%%%%%%%%%%%%%%%%%%%%%%%
\section{p-shell nuclei}
 \label{Sec:nuclei}
We now turn to heavier $p$-shell nuclei. For simplicity,
we ignore the proton-neutron mass difference, and use the same nucleon
mass $m=938.92$~MeV for the protons and the neutrons; furthermore, we
add a standard repulsive Coulomb potential between the protons.

We use the No-Core Configuration Interaction (NCCI)
approach~\cite{Barrett:2013nh} to determine the ground states of
$p$-shell nuclei (excluding mirror nuclei) at N$^2$LO; for select
nuclei we perform calculations at lower orders, and include narrow
low-lying excited states as well.  In the NCCI approach we expand the
wave function $\Psi$ of a nucleus consisting of $A$ nucleons in an
$A$-body basis of Slater determinants $\Phi_k$ of single-particle wave
functions $\phi_{nljm}(\vec{r})$.  Here, $n$ is the radial quantum
number, $l$ the orbital motion, $j$ the total spin from orbital motion
coupled to the intrinsic nucleon spin, and $m$ the spin-projection.
The Hamiltonian ${\hat H}$ is also expressed in this basis and
thus the many-body Schr\"odinger equation becomes a matrix eigenvalue
problem; for an NN potential plus 3NFs, this matrix is sparse for $A >
4$.  The eigenvalues of this matrix are approximations to the energy
levels, to be compared to the experimental binding energies and
spectra, and the corresponding eigenvectors to the nuclear wave
functions.

We use the conventional harmonic oscillator (HO) basis with energy
parameter $\hbar\omega$ for the single-particle wave functions, in
combination with a truncation on the total number of HO quanta in the
system: the basis is limited to many-body basis states with $\sum_{A}
N_i \le N_0 + \nmax$, with $N_0$ the minimal number of quanta for that
nucleus and \nmax\ the truncation parameter.  (Even/odd values of
\nmax\ provide results for natural/unnatural parity.)  Numerical
convergence toward the exact results for a given Hamiltonian is
obtained with increasing \nmax, and is marked by approximate
\nmax\ and \hw\ independence.  In practice, we use extrapolations to
estimate the binding energy in the complete (but infinite-dimensional)
space~\cite{Maris:2008ax,Coon:2012ab,Furnstahl:2012qg,More:2013rma,Wendt:2015nba},
based on a series of calculations in finite bases.

The rate of convergence depends both on the nucleus and on the
interaction.  For realistic interactions, the dimension of the matrix
needed to reach a sufficient level of convergence is in the tens or
even hundreds of billions, which saturates or exceeds the capabilities
of current high-performance computing facilities.  In order to improve
the convergence of the basis space expansion, we therefore first apply
a Similarity Renormalization Group (SRG)
transformation~\cite{Bogner:2007rx,Bogner:2009bt,Roth:2013fqa} to
soften these interactions.  Most of the results presented here have
been evolved to an SRG parameter of $\alpha=0.08$~fm$^4$ and all of them include any
induced 3NFs, but we have also performed calculations at
$\alpha=0.04$~fm$^4$ in order to make sure the dependence on the SRG
parameter is weak.  For $^4$He we also compare with the Yakubovsky
calculations presented in the previous section.

The calculations described in this section have been performed with
the NCCI code MFDn~\cite{doi:10.1002/cpe.3129,SHAO20181} to calculate
the lowest energy levels with natural parity of $p$-shell nuclei.
Most NCCI calculations were performed on the IBM BG/Q Mira at the
Argonne Leadership Computing Facility (ALCF), with additional
calculations performed on the Cray XC40 Theta at ALCF and the Cray
XC40 Cori at the National Energy Scientific Computing Center (NERSC).
For all nuclei we have performed calculations for at least four
\hw\ values (one below, and two above the variational minimum), in
order to perform extrapolations to the complete (infinite-dimensional)
basis with uncertainty estimates.  We use a simple 3-point exponential
extrapolation in \nmax\ at fixed \hw,
\begin{eqnarray}
  E^{\hw}(\nmax) &=& E^{\hw}_\infty + a \, {\rm e}^{(-b \, \nmax)}
  \;,
\end{eqnarray}
using three consecutive values of \nmax\ around the variational
minimum in \hw, which seems to work well for a range of interactions
and nuclei~\cite{Maris:2008ax,Maris:2013poa,Jurgenson:2013yya}.  We
take as our best estimate for $E_\infty$ in the complete basis the
value of $E^{\hw}_\infty$ for which $|E^{\hw}_\infty - E^{\hw}(\nmax)|$
is minimal.  Our estimate of the extrapolation uncertainty is given by
the maximum of
\begin{itemize}
\item the difference in $E^{\hw}_\infty$ for two successive
  extrapolations using data for $(\nmax-6, \nmax-4, \nmax-2)$ and
  $(\nmax-4, \nmax-2, \nmax)$ respectively;
\item half the variation in $E^{\hw}_\infty$ over a 8~MeV interval in
  \hw\ around the variational minimum;
\item 20\% of $|E^{\hw}_\infty - E^{\hw}(\nmax)|$.
\end{itemize}
This procedure is identical to what was used in
Refs.~\cite{Binder:2018pgl,Epelbaum:2018ogq}.
When we apply this method to $^3$H, our results are, 
within the estimated extrapolation uncertainties, in excellent agreement 
with the Yakubosky results discussed in the previous section, see Table~\ref{tab:3H3He}.
As expected, the NCCI calculations performed after SRG evolution to $\alpha=0.08$~fm$^4$ converge faster, with estimated extrapolation uncertainties in the ground state energies of about ten keV,
whereas the extrapolation uncertainties are up to about a hundred keV at $\alpha=0.04$~fm$^4$.
Within these uncertainties, our results are also independent of the SRG parameter $\alpha$.

In Table~\ref{Tab:res_A4A6} we list our order-by-order results for
the ground state energies of $^4$He and $A=6$ nuclei for $\alpha=0.04$
and $0.08$~fm$^4$.  The quoted uncertainties in
Table~\ref{Tab:res_A4A6} are the extrapolation uncertainties only.
For $^4$He we can perform these calculations up to $\nmax=14$, which
is sufficient to obtain the ground state energies to within a few keV
for both $\alpha=0.04$ and $0.08$~fm$^4$.  We do observe a small dependence on the SRG parameter, of the order of a few tens of keV, 
which can easily be caused by induced 4-body forces from the SRG evolution which have not been incorporated here.  Indeed, compared to the results from the Yakubovsky calculations discussed in the previous section, we see differences of up to about a hundred keV in the binding energies (i.e. up to 0.4\%), with the Yakubovsky results being slightly deeper bound, see Table~\ref{tab:4He}.  This suggests that the missing induced 4-body forces would lead to slightly stronger binding, at least in $^4$He.

For $^6$He and $^6$Li our calculations are limited by the number of the
input 3NF matrix elements, which in practice means that we can perform
calculations up to $\nmax=12$.  Clearly, our results are not as
precise as for $^4$He, and in fact, our NCCI extrapolation
uncertainties at $\alpha=0.04$~fm$^{4}$ are of the same order of
magnitude as the difference between the $\alpha=0.04$ and
$\alpha=0.08$~fm$^{4}$ results, whereas the extrapolation
uncertainties at $\alpha=0.08$~fm$^{4}$ are significantly smaller, again as
expected.  We therefore concentrate on our results with
$\alpha=0.08$~fm$^{4}$ for nuclei with $A > 6$.

Comparing the results at different chiral orders, we see that there is,
not surprisingly, a large difference between the LO and the NLO results, 
followed by a significantly smaller difference between the
NLO and the N$^2$LO predictions.  The role of the 3NFs at N$^2$LO is
significant, and in fact, while the NN-only potential at N$^2$LO
decreases the binding energies by about 1 to 2 MeV compared to the NLO
potential, which moves the ground state energies of $^6$He and $^6$Li
further away from experiment, with the complete N$^2$LO interaction
including the 3NFs the ground state energies of both $^6$He and $^6$Li
are within a few hundred keV of the experimental values.

%%%%%%%%%%%%%%%%%%%%%%%%%%%%%%%%%%%%%%%%%%%%%%%%%%%%%%%%%%%%%%%%%%%%%%%%%%%%%
\begin{figure*}[bth]
  \includegraphics[width=0.9\columnwidth]{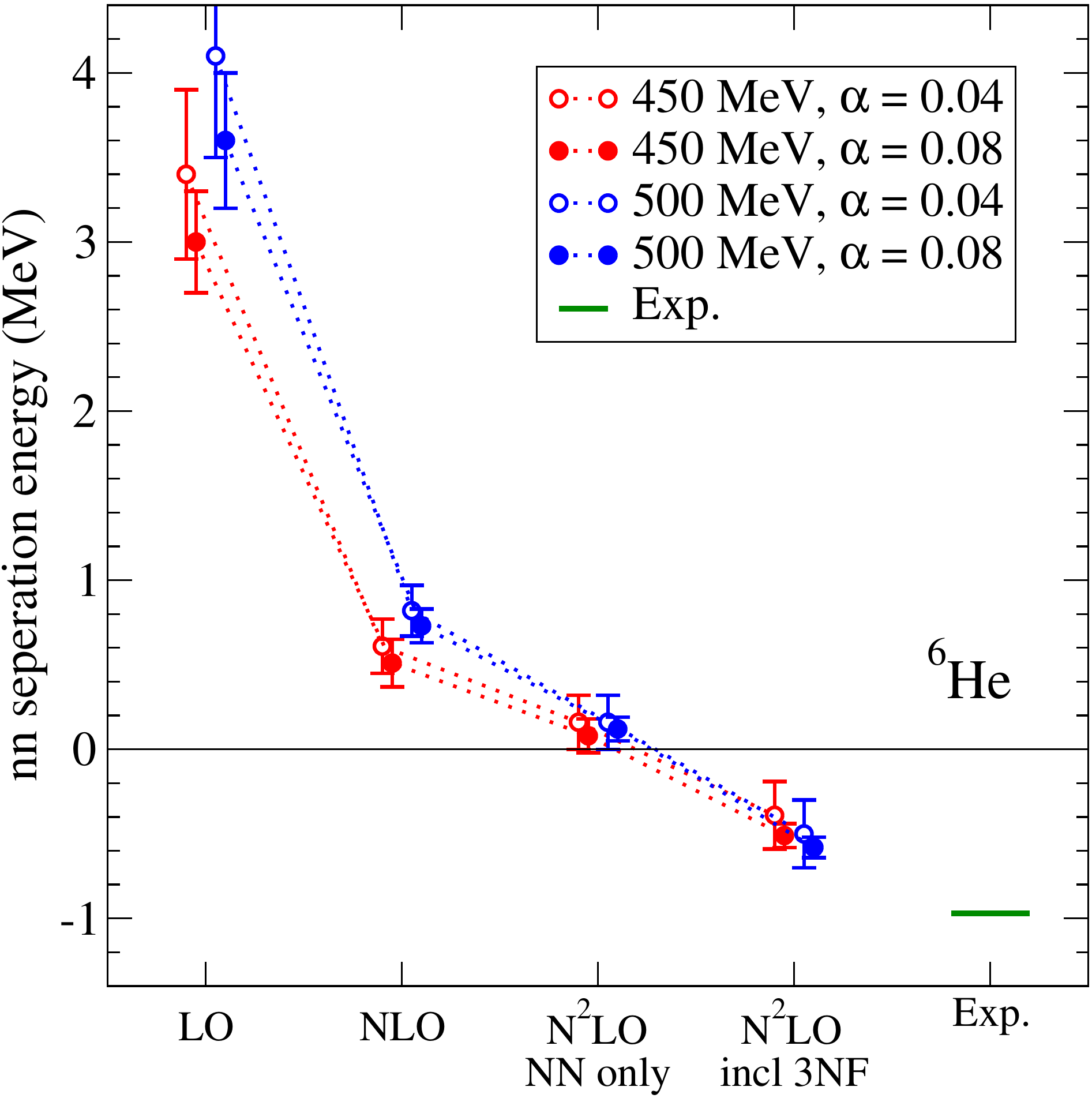} \qquad
  \includegraphics[width=0.9\columnwidth]{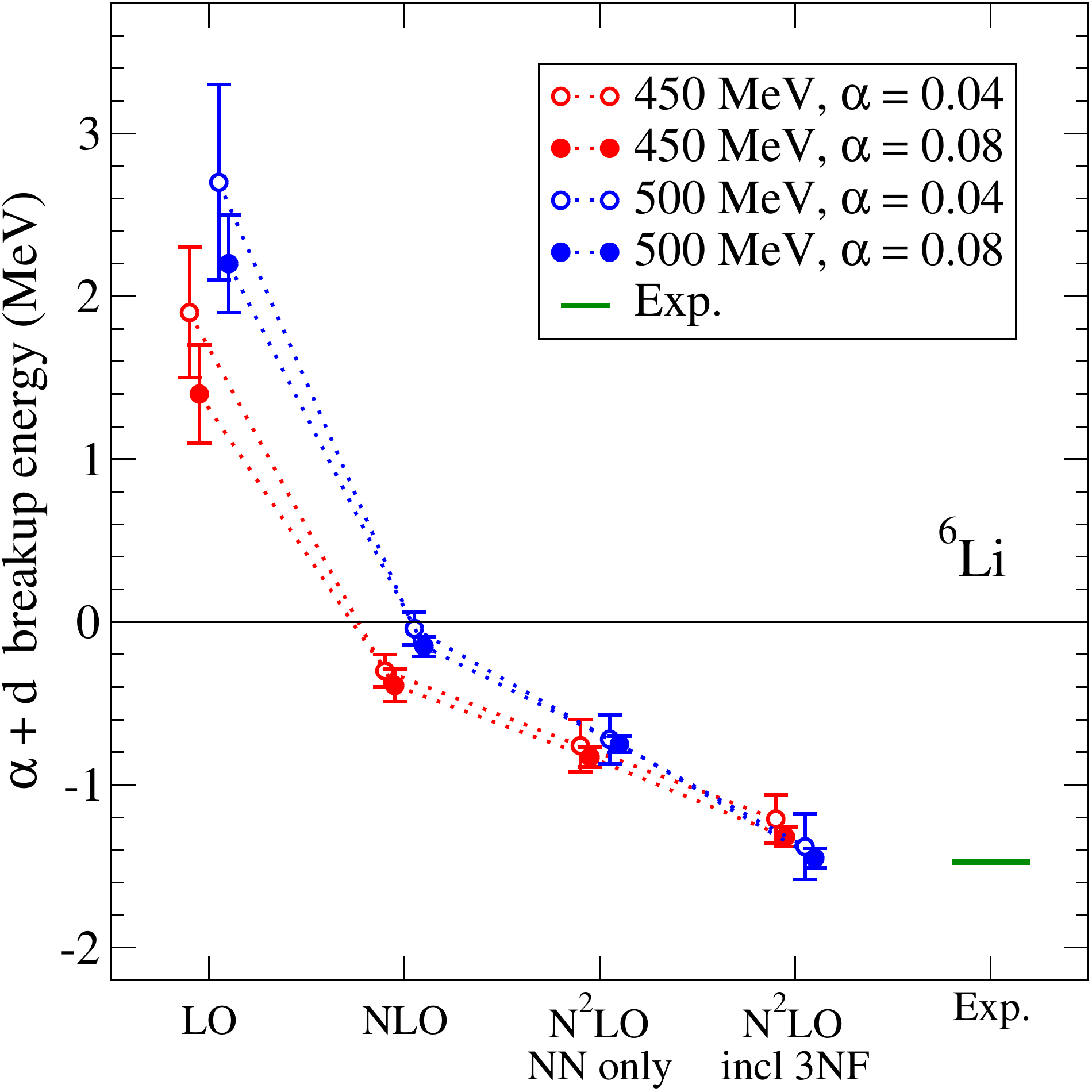}
  \caption{\label{Fig:res_6He6Li} (Color online)
    Order-by-order $^6$He $nn$ separation energy (left)
    and $^6$Li $\alpha + d$ breakup threshold (right).
    Error bars indicate the NCCI extrapolation uncertainties only.}
\end{figure*}
%%%%%%%%%%%%%%%%%%%%%%%%%%%%%%%%%%%%%%%%%%%%%%%%%%%%%%%%%%%%%%%%%%%%%%%%%%%%%

Our results indicate that at LO, neither $^6$Li nor $^6$He is bound,
as is illustrated in Fig.~\ref{Fig:res_6He6Li}.  Here, we calculate
the $nn$ separation energy and the $\alpha+d$ breakup threshold as the
difference between the extrapolated energies of $^6$He and $^4$He, and
between that of $^6$Li and $E_\alpha+E_d$, respectively; for the
numerical uncertainty we take the extrapolation uncertainty of the
$A=6$ nucleus.  It turns out that $^6$He only becomes bound at N$^2$LO
including 3NFs, both with the $450$~MeV and the $500$~MeV regulator.
And even then it is bound relative to $^4$He by only $0.5$ to
$0.6$~MeV, whereas experimentally it is bound by about 1 MeV.  On the
other hand, $^6$Li becomes minimally bound at NLO, and its binding
relative to the $\alpha + d$ threshold is in good agreement with the
experimental breakup threshold for both regulator values.  Note that
this is qualitatively similar to what we found in
Ref.~\cite{Epelbaum:2018ogq} using the SCS interactions.

In addition to the ground state energies, we also list in
Table~\ref{Tab:res_A4A6} the excitation energies of the lowest excited
states in $^6$He and $^6$Li at $\alpha=0.08$~fm$^{4}$.  For these
excitation energies, we extrapolate the total energy of these excited
states to the complete (infinite-dimensional) basis, following the
same procedure as for the ground states.  Next, we take the difference
between these extrapolated energies as our best estimate for the
excitation energy.  For the corresponding extrapolation uncertainties
we take the maximum of the estimated extrapolation uncertainties of
the total energies of the two states, which is a rather conservative
uncertainty estimate given the often strong correlations between
different states in the spectrum.  For narrow excited states (and we
mainly consider narrow excited states in this work), this
extrapolation method seems to give results that are numerically
reasonably stable, even for states that are above threshold, and has
the advantage that we can apply the same method to all nuclei under
consideration.  Our results clearly indicate (not surprisingly!) that
there are strong correlations between the ground state and the excited
state -- the difference between the excitation energies at different
chiral orders is significantly smaller than the difference between the
ground state energies at different chiral orders.  We will come back
to this when we discuss the uncertainties associated with the
truncation of the chiral expansion in the next section.  At
N$^2$LO, including consistent 3NFs, the excitation energies are within
a few hundred keV of the experimental values, which is similar to the
deviation of the total ground state energies from their experimental
values.  (Our results at $\alpha=0.04$~fm$^{4}$ are within the quoted
extrapolation uncertainty estimates, but with larger numerical
uncertainties.)

%%%%%%%%%%%%%%%%%%%%%%%%%%%%%%%%%%%%%%%%%%%%%%%%%%%%%%%%%%%%%%%%%%%%%%%%%%%%%
\begin{table*}
   \begin{ruledtabular}
    \begin{tabular*}{\textwidth}{@{\extracolsep{\fill}}lc|dddd|d}
     %      \begin{tabular}{lc|dddd|d}
 &&&&&&\\[-12pt]     
    $^A$Z$(J^{\pi},T)$ & $\Lambda$ 
                       & \multicolumn{1}{c}{LO}
                       & \multicolumn{1}{c}{NLO}
                       & \multicolumn{1}{c}{N$^2$LO, NN-only}
                       & \multicolumn{1}{c|}{N$^2$LO NN+3NFs}
                       & \multicolumn{1}{c}{Exp. (MeV)} \\
    \hline
    $^7$Li$(\frac{3}{2}^-,\frac{1}{2})$
                       & 450 &   -61.35(15)&  -38.72(9)  &  -36.98(11) &  -39.39(6)  & -39.24 \\
                       & 500 &   -62.1(2)  &  -36.82(11) &  -35.10(14) &  -39.73(6)  & \\
    $E_x(\frac{1}{2}-,\frac{1}{2})$
                       & 450 &    -0.07(13)&  0.22(11) &    0.20(12) &    0.32(7)  &   0.48 \\
                       & 500 &    -0.09(14)&  0.17(12) &    0.16(15) &    0.35(6)  & \\
    $E_x(\frac{7}{2}-,\frac{1}{2})$
                       & 450 &    10.22(45)&  5.54(18) &    5.30(23) &    4.86(9)  & 4.63 \\
                       & 500 &    10.49(60)&  5.40(13) &    5.05(21) &    4.81(8)  & \\
    $E_x(\frac{5}{2}-,\frac{1}{2})$
                       & 450 &    10.10(43)&    6.70(16) &    6.50(17) &    6.65(12) & 6.68(5) \\
                       & 500 &    10.37(57)&    6.37(17) &    6.16(18) &    6.77(10) & \\
    $E_x(\frac{5}{2}-,\frac{1}{2})$
                       & 450 &    16.55(50)&  8.64(25) &    8.16(16) &    7.84(11) & 7.46 \\
                       & 500 &    17.3(6)  &  8.37(16) &    7.73(16) &    7.79(9)  &  \\
    \hline
    $^8$He$(0^+, 2)$   & 450 &   -41.6(9)  &  -28.2(7)   &  -27.1(5)   &  -30.4(2)   & -31.41 \\
                       & 500 &   -41.6(1.0)&  -26.3(6)   &  -25.6(4)   &  -30.9(2)   & \\
    \hline
    $^8$Li$(2^+, 1)$   & 450 &   -59.5(3)  &  -39.44(19) &  -38.07(19) &  -41.23(16) & -41.28 \\
                       & 500 &   -59.6(4)  &  -37.24(14) &  -36.11(16) &  -41.85(15) & \\
    $E_x(1^+, 1)$      & 450 &     0.50(27)&  0.90(20) &    0.80(23) &    1.11(19) &   0.98 \\
                       & 500 &     0.49(41)&  0.80(16) &    0.72(17) &    1.10(18) & \\
    $E_x(3^+, 1)$      & 450 &     4.57(39)&  3.00(28) &    2.82(27) &    2.54(15) &   2.26 \\
                       & 500 &     4.49(48)&  2.92(20) &    2.68(22) &    2.42(15) & \\
    $E_x(0^+, 1)$      & 450 &    -0.71(23)&  2.01(22) &    2.15(28) &    3.04(26) &   ? \\
                       & 500 &    -0.74(26)&  2.26(24) &    2.04(26) &    3.39(27) &  \\
    $E_x(4^+, 1)$      & 450 &    10.65(35)&  6.52(35) &    6.36(33) &    6.80(24) &   6.54(2)\\
                       & 500 &    11.03(45)&  6.24(30) &    6.05(27) &    6.90(23) & \\
    \hline
    $^{10}$Be$(0^+, 1)$& 450 &   -97.7(1.5)&   -61.9(6)  &  -60.8(4)   &  -66.5(5)   & -64.98 \\
                       & 500 &   -98.1(1.7)&   -57.9(6)  &  -57.3(5)   &  -67.5(4)   & \\
    $E_x(2^+, 1)$      & 450 &     7.6(2.1)&   3.5(8)  &    3.3(5)   &    3.3(6)   &   3.37 \\
                       & 500 &     8.1(2.5)&   3.4(7)  &    3.0(5)   &    3.2(4)   & \\
    $E_x(2^+, 1)$      & 450 &     6.1(1.6)&   4.6(9)  &    4.8(6)   &    6.3(7)   &   5.96 \\
                       & 500 &     6.6(2.0)&   4.2(7)  &    4.4(6)   &    6.2(6)   & \\
    \hline
    $^{10}$B$(3^+, 0)$ & 450 &   -92.8(1.6)&   -61.1(6)  &  -60.3(4)   &  -66.4(4)   & -64.75 \\
                       & 500 &   -92.5(2.0)&   -57.0(5)  &  -57.0(5)   &  -68.4(4)   & \\
    $E_x(1^+, 0)$      & 450 &     0.2(1.7)&   1.8(8)  &    1.6(5)   &    1.4(6)   &   0.72 \\
                       & 500 &     0.2(2.0)&   1.5(7)  &    1.5(5)   &    1.8(5)   & \\
    $E_x(1^+, 0)$      & 450 &    -6.7(1.6)&    -1.4(8)  &   -0.8(5)   &    1.7(1.0) &   2.15 \\
                       & 500 &    -7.0(2.0)&  -1.6(6)  &   -0.8(5)   &    2.2(5)   & \\
    $E_x(2^+, 0)$      & 450 &    -0.6(1.9)&   2.1(6)  &    2.2(5)   &    3.4(5)   &   3.59 \\
                       & 500 &    -0.8(2.1)&   1.6(5)  &    1.9(5)   &    4.1(5)   & \\
    $E_x(3^+, 0)$      & 450 &     1.5(2.2)&   3.8(1.1)&    4.2(7)   &    5.7(7)   &   4.77 \\
                       & 500 &     1.3(2.6)&   3.3(9)  &    3.9(5)   &    7.1(6)   & \\
    \hline
    $^{12}$B$(1^+, 1)$ & 450 &  -113.7(1.3)&   -76.0(7)  &   -76.7(5)  &  -84.8(4)   & -79.58 \\
                       & 500 &  -111.7(1.6)&   -70.4(6)  &   -72.6(5)  &  -87.5(4)   & \\
    $E_x(2^+, 1)$      & 450 &     4.4(1.3)&     1.2(8)  &     0.7(5)  &   -0.9(4)   &   0.95 \\
                       & 500 &     4.6(1.7)&     1.4(7)  &     0.6(5)  &   -1.1(4)   & \\
    $E_x(0^+, 1)$      & 450 &    -1.3(1.3)&     0.3(9)  &     0.7(5)  &    1.9(6)   &   2.72 \\
                       & 500 &    -1.4(1.6)&     0.1(8)  &     0.5(5)  &    2.7(6)   & \\
    $E_x(2^+, 1)$      & 450 &     0.0(1.4)&     1.8(9)  &     2.0(6)  &    3.4(7)   &   3.76 \\
                       & 500 &     0.0(1.7)&     1.5(8)  &     1.8(6)  &    4.1(6)   & \\
    $E_x(1^+, 1)$      & 450 &     2.1(1.6)&     3.0(8)  &     3.2(5)  &    4.9(7)   &   4.99 \\
                       & 500 &     2.3(2.0)&     2.6(6)  &     2.9(6)  &    5.7(6)   & \\
    $E_x(3^+, 1)$      & 450 &     4.9(1.4)&     3.8(9)  &     4.1(5)  &    5.3(7)   &   5.61 \\
                       & 500 &     5.2(1.8)&     3.6(8)  &     3.8(5)  &    6.1(7)   & \\
    \hline
    $^{12}$C$(0^+, 0)$ & 450 &  -145.0(1.0)&  -89.7(5)   &   -90.0(5)  &  -98.7(4)   & -92.16 \\
                       & 500 &  -144.6(1.2)&  -83.3(5)   &   -85.0(4)  & -101.8(4)   & \\ 
    $E_x(2^+, 0)$      & 450 &     6.9(0.9)&    3.4(6)   &     3.2(4)  &    4.2(4)   &   4.44 \\
                       & 500 &     7.5(1.3)&    3.1(6)   &     2.9(5)  &    4.5(4)   & \\
    $E_x(1^+, 0)$      & 450 &    31.3(1.2)&   14.2(6)   &    12.9(5)  &    9.6(4)   &  12.71 \\
                       & 500 &    32.2(1.4)&   13.6(6)   &    11.8(5)  &    9.9(4)   & \\
    $E_x(4^+, 0)$      & 450 &    23.3(1.1)&   12.2(7)   &    11.7(5)  &   13.7(5)   &  14.08 \\ 
                       & 500 &    24.6(2.0)&   11.4(6)   &    10.8(5)  &   14.6(4)   & \\[-2pt]
%    \hline
    %
  \end{tabular*}
  \caption{\label{Tab:res_A7A8A10A12}
    Order-by-order results for ground state energies of select stable
    $p$-shell nuclei, with excitation energies for narrow states with
    natural parity, from LO up through N$^2$LO including 3NFs.
    Energies and cutoffs are in MeV.
    Calculated results shown are for $\alpha = 0.08$~fm$^{4}$, with the NCCI 
    extrapolation uncertainties only.
    Experimental values are from Refs.~\cite{Wang_2017,TILLEY20023,TILLEY2004155,KELLEY201771}. 
    }
\end{ruledtabular}   
\end{table*}
%%%%%%%%%%%%%%%%%%%%%%%%%%%%%%%%%%%%%%%%%%%%%%%%%%%%%%%%%%%%%%%%%%%%%%%%%%%%%

In Table~\ref{Tab:res_A7A8A10A12} we summarize our results for the
ground state energies of a range of $p$-shell nuclei, as well as the
excitation energies for narrow excited states with natural parity
(i.e. positive for even nuclei, and negative for odd nuclei in the
$p$-shell).  We also include the incomplete results at N$^2$LO
without the 3NFs in order to highlight the importance of the 3NFs
starting at this chiral order.  The overall convergence pattern of the
ground state binding energies starts out similar to that of $^4$He,
$^6$He, and $^6$Li: significant overbinding at LO, and modest
underbinding at NLO.  Furthermore, it is interesting to note that for all
ground state energies in Table~\ref{Tab:res_A7A8A10A12}, the difference 
between N$^2$LO calculations with or without 3NFs is noticeably larger
than the difference between NLO and N$^2$LO calculations without 3NFs.  
This highlights the importance of 3NFs at N$^2$LO.

Looking in more detail at specific nuclei up to $A=10$, the N$^2$LO NN-only potential leads to even more underbinding than at NLO, but the 3NFs at N$^2$LO increase the
binding energy again.  The additional binding coming from the 3NFs is
significantly stronger with $\Lambda=500$~MeV than with
$\Lambda=450$~MeV, and overbinds $^7$Li slightly compared to
experiment.  Note that for $A=8$ however, both regulators underbind $^8$He 
at N$^2$LO including 3NFs, while the ground state energy of $^8$Li 
is in agreement with experiment for $\Lambda=450$~MeV (to within the extrapolation uncertainty), and slightly overbound for $\Lambda=500$~MeV.  The amount of overbinding keeps increasing with $A$;
at $A=10$ the $\Lambda=450$~MeV interaction also leads to overbinding.  

\begin{figure}[tb]
  \includegraphics[width=0.9\columnwidth]{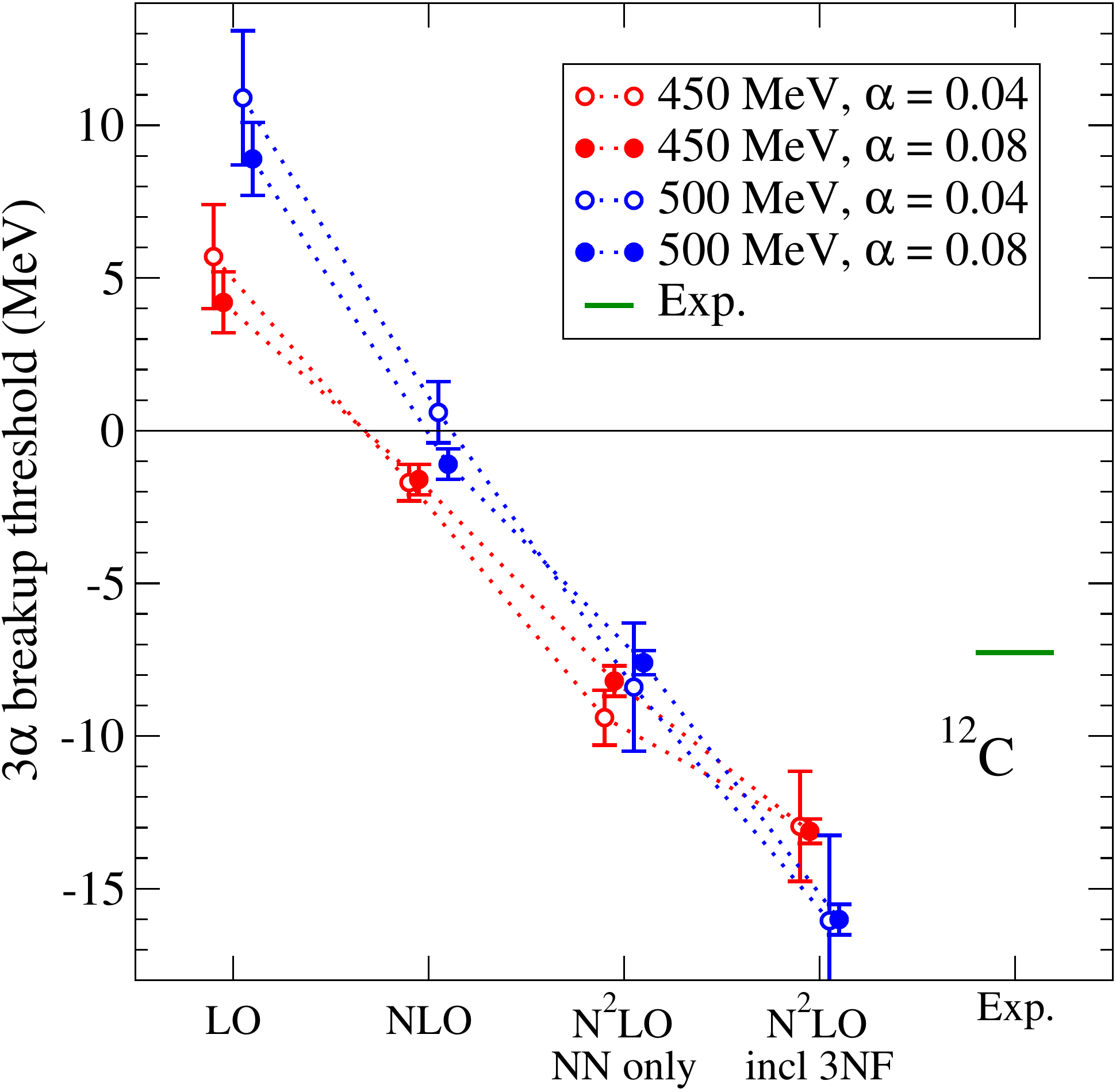}
  \caption{\label{Fig:res_12C3alpha}
    (Color online) Order-by-order $3\,\alpha$ breakup threshold of $^{12}$C.  Error bars indicate the NCCI extrapolation uncertainties only.}
\end{figure}

For $A=12$ the N$^2$LO NN-only potential increases the binding
slightly compared to NLO, in contrast to the binding energies up to $A=10$.  Adding the 3NFs leads to significant overbinding for $^{12}$C.  The overbinding also means that the lowest breakup thresholds appear higher in the spectrum than in reality.  In particular, the $3\alpha$ threshold in $^{12}$C is around 15~MeV at N$^2$LO, that is, about a factor of 2 higher than in the real world, as is illustrated in Fig.~\ref{Fig:res_12C3alpha}. 
%%%%%%%%%%%%%%%%%%%%%%%%%%%%%%%%%%%%%%%%%%%%%%%%%%%%%%%%%%%%%%%%%%%%%%%%%%%%%
%%%%%%%%%%%%%%%%%%%%%%%%%%%%%%%%%%%%%%%%%%%%%%%%%%%%%%%%%%%%%%%%%%%%%%%%%%%%
One would therefore anticipate that the first excited $0^+$ in $^{12}$C 
(also known as the Hoyle state), which is experimentally near the $3\alpha$ threshold, 
will be significantly too high in the spectrum at N$^2$LO.  Unfortunately, 
within the NCCI approach we cannot correctly describe this state within the numerically accessible basis spaces~\cite{Neff:2012es}.

%%%%%%%%%%%%%%%%%%%%%%%%%%%%%%%%%%%%%%%%%%%%%%%%%%%%%%%%%%%%%%%%%%%%%%%%%%%%%
\begin{figure}[tb]
  \includegraphics[width=\columnwidth]{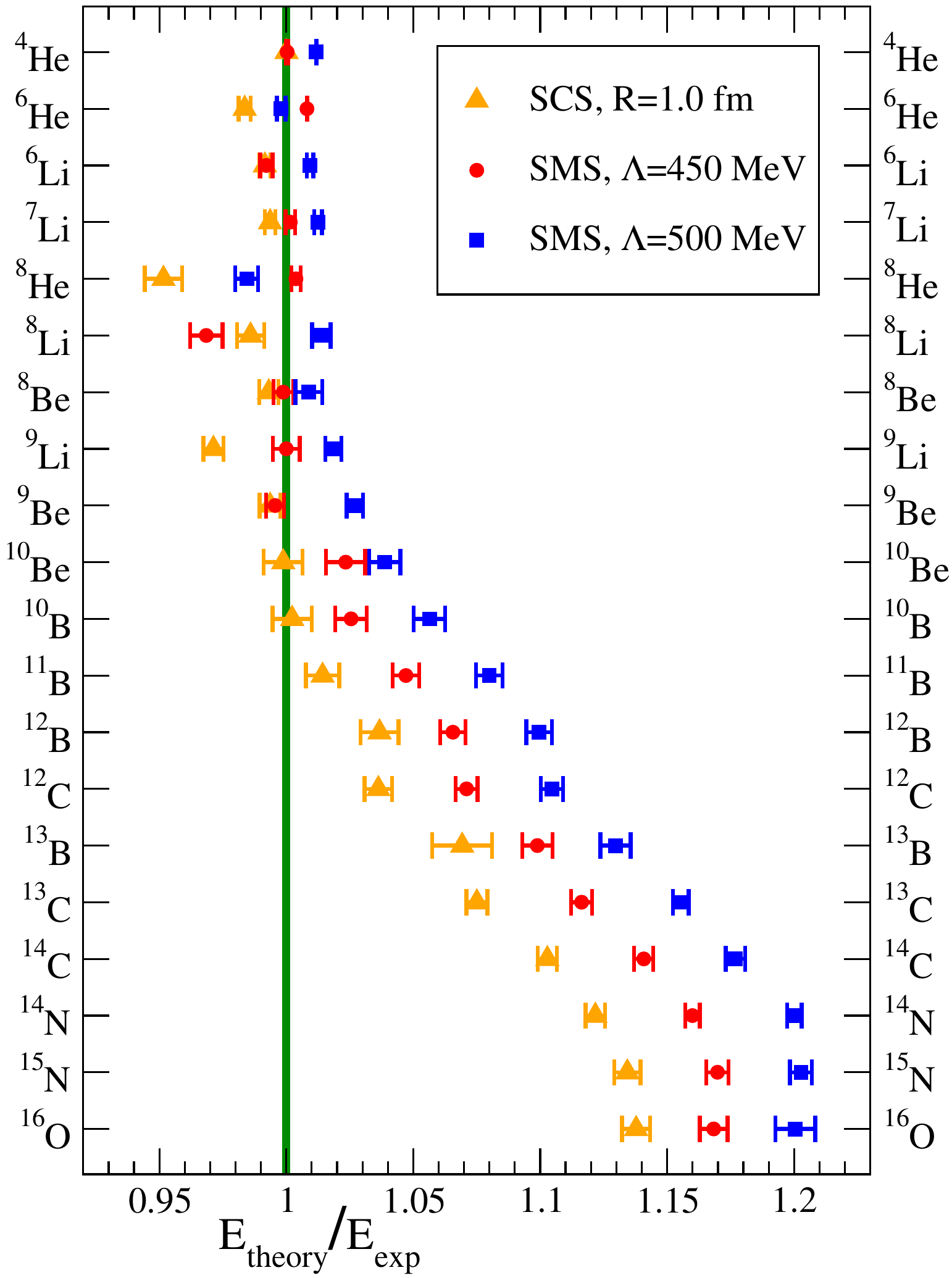}
  \caption{\label{Fig:Egs_comparison}
    (Color online) Comparison of ground state energies of $p$-shell nuclei between chiral EFT calculations at N$^2$LO and experiment.
    All results shown are for $\alpha = 0.08$~fm$^{4}$ and
    error bars indicate the NCCI extrapolation uncertainties only.}
\end{figure}
%%%%%%%%%%%%%%%%%%%%%%%%%%%%%%%%%%%%%%%%%%%%%%%%%%%%%%%%%%%%%%%%%%%%%%%%%%%%%
\begin{table}
  \begin{ruledtabular}
    \begin{tabular*}{\textwidth}{@{\extracolsep{\fill}}l|d|d|d}
     %      \begin{tabular}{l|d|d|d}
 &&&\\[-12pt]     
    $^A$Z$(J^{\pi},T)$ & \multicolumn{1}{c|}{N$^2$LO(450)} 
                       & \multicolumn{1}{c|}{N$^2$LO(500)}
                       & \multicolumn{1}{c}{Exp. (MeV)} \\
    \hline
    $^4$He$(0^+, 0)$               & -28.527(2) & -28.630(2) & -28.296 \\
                                   &  +0.231(2) &  +0.334(2) &         \\ \hline
    $^6$He$(0^+, 1)$                 & -29.04(7) & -29.21(6) & -29.27 \\
                                     &  -0.23(7) & -0.06(6) &         \\ \hline
    $^6$Li$(1^+, 0)$                 & -32.04(6) & -32.29(4) & -31.99 \\
                                     &  +0.05(6) &  +0.30(4) &         \\ \hline
    $^7$Li$(\frac{3}{2}^-,\frac{1}{2})$& -39.39(6)& -39.73(6)& -39.24 \\
                                     &    +0.15(6) & +0.49(6)&         \\ \hline
    $^8$He$(0^+, 1)$                 & -30.4(2)  & -30.9(2)  & -31.41 \\
                                     &  -1.0(2)  &  -0.5(2)  &         \\ \hline
    $^8$Li$(2^+, 0)$                 & -41.23(16)& -41.85(15)& -41.28 \\
                                     &  -0.05(16)&  +0.57(15)&         \\ \hline
    $^8$Be$(0^+, 0)$                 & -56.5(3)  & -57.0(3)  & -56.50 \\
                                     &   0.0(3)  &  +0.5(3)  &         \\ \hline
    $^9$Li$(\frac{3}{2}^-,\frac{3}{2})$&-45.14(16)&-46.18(16)& -45.34 \\
                                      &  -0.20(16)& +0.84(16)&         \\ \hline
    $^9$Be$(\frac{3}{2}^-,\frac{1}{2})$&-58.82(21)&-59.73(16)& -58.16 \\
                                      &  +0.66(21)& +1.57(16)&         \\ \hline
    $^{10}$Be$(0^+, 1)$               & -66.5(5)  & -67.5(4) & -64.98 \\
                                      &  +1.5(5)  &  +1.5(4)           \\ \hline
    $^{10}$B$(3^+, 0)$                & -66.4(4)  & -68.4(4) & -64.75 \\
                                      &  +1.7(4)  &  +3.7(4) &         \\ \hline
    $^{11}$B$(\frac{3}{2}^-,\frac{1}{2})$&-79.8(4)& -82.3(4) & -76.21 \\ 
                                      &    +3.6(4)&  +6.1(4) &         \\ \hline
    $^{12}$B$(1^+, 1)$                & -84.8(4)  & -87.5(4) & -79.58 \\
                                      &  +5.2(4)  &  +7.9(4) &         \\ \hline
    $^{12}$C$(0^+, 0)$                & -98.7(4)  &-101.8(4) & -92.16 \\
                                      &  +6.5(4)  &  +9.6(5) &         \\ \hline
    $^{13}$B$(\frac{3}{2}^-,\frac{3}{2})$& -92.8(5)& -95.4(5)& -84.45 \\
                                         &  +8.4(5)& +11.0(5)&         \\ \hline
    $^{13}$C$(\frac{1}{2}^-,\frac{1}{2})$&-108.3(4)&-112.2(4)& -97.11 \\
                                         & +11.2(4)& +15.1(4)&         \\ \hline
    $^{14}$C$(0^+, 1)$                  &-120.1(4) &-123.9(4) & -105.28 \\
                                        & +14.8(4) & +18.6(4) &         \\ \hline
    $^{14}$N$(1^+, 0)$                  &-121.4(4) &-125.6(4) & -104.66 \\
                                        & +16.7(4) & +20.9(4) &         \\ \hline
    $^{15}$N$(\frac{1}{2}^-,\frac{1}{2})$&-135.1(5)&-138.9(5) & -115.49 \\
                                         & +19.6(5)& +23.4(5) &         \\ \hline
    $^{16}$O$(0^+, 0)$                  &-149.1(7) &-153.2(1.0)& -127.62 \\
                                        & +21.5(7) & +25.6(1.0)&         \\ [-2pt]
  \end{tabular*}
  \caption{\label{Tab:res_Egs_summary}
    ground state energies of $p$-shell nuclei, excluding mirror nuclei,
    at N$^2$LO including 3NFs with SRG $\alpha = 0.08$~fm$^{4}$, 
    together with the deviations from the experimental values. Quoted uncertainties are the extrapolation uncertainties only. 
    Energies and cutoffs are in MeV. Experimental values are from Ref.~\cite{Wang_2017}.}
 \end{ruledtabular}    
\end{table}
%%%%%%%%%%%%%%%%%%%%%%%%%%%%%%%%%%%%%%%%%%%%%%%%%%%%%%%%%%%%%%%%%%%%%%%%%%%%%
%

As we continue up in the $p$-shell and move to $^{16}$O, the
amount of overbinding at N$^2$LO gets even worse, as is illustrated in
Fig.~\ref{Fig:Egs_comparison}.  For comparison, we also include our
results with the SCS~\cite{Epelbaum:2018ogq} interaction at N$^2$LO in
this figure, which for the considered cutoff values agrees somewhat better with experiment in the upper half of the $p$-shell.  The calculated ground state energies, together with the corresponding experimental values, as well as the
deviation from these experimental values, are summarized in
Table~\ref{Tab:res_Egs_summary}.

\begin{figure}[tb]
  \includegraphics[width=\columnwidth]{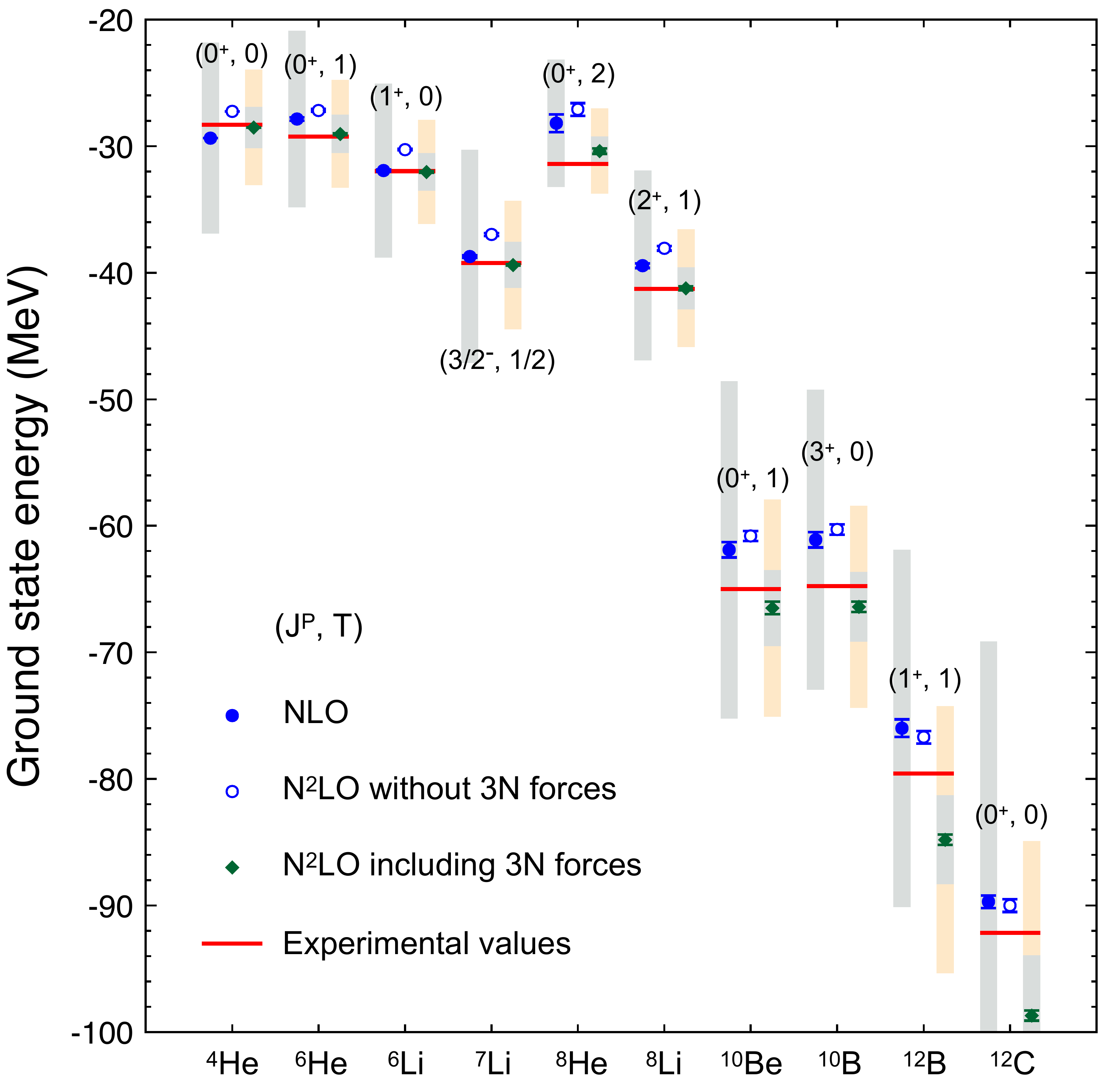}
  \caption{\label{Fig:GSEnergies} (Color online)
Calculated ground state energies in MeV using chiral NLO, and N$^2$LO
interactions at $\Lambda = 450$~MeV (blue and green symbols) in comparison with
experimental values (red levels). For each nucleus the NLO, and
N$^2$LO results are the left and right symbols and bars,
respectively.
The open blue symbols correspond to incomplete calculations at N$^2$LO
using NN-only interactions. Blue error bars indicate the NCCI
extrapolation uncertainty. All results shown are for $\alpha =
0.08$~fm$^{4}$.  The light (coral) and dark (grey) shaded bars indicate the $95\%$
and $68\%$ DoB truncation errors, respectively, estimated using the Bayesian model $\bar
C_{0.5-10}^{650}$. }
\end{figure}

Next, we show in Fig.~\ref{Fig:GSEnergies}
%%%%%%%%%%%%%%%%%%%%%%%%%%%%%%%%%%%%%%%%%%%%%%%%%%%%%%%%%%%%%%%%%%%%%%%%%%%%%
%%%%%%%%%%%%%%%%%%%%%%%%%%%%%%%%%%%%%%%%%%%%%%%%%%%%%%%%%%%%%%%%%%%%%%%%%%%%
the calculated ground state energies together with the corresponding
truncation errors for those nuclei for which the results are available
at all orders up through N$^2$LO. Here, we use the Bayesian truncation
model $\bar C_{0.5-10}^{650}$ and assume the expansion parameter $Q
= M_\pi^{\rm eff} / \Lambda_b = 200/650 \sim 0.31$. We, however,
emphasize that such an estimation is somewhat simplistic, see
Ref.~\cite{Binder:2018pgl} and the discussion in the next
section. Also, 
the assumed value for the expansion parameter may be too optimistic for heavier nuclei. 
This is e.g.~indicated by the spread between the (obviously correlated)
predictions for the ground state energies for $\Lambda = 450$ and $500$~MeV in
Fig.~\ref{Fig:Egs_comparison}, which can serve as a measure of the
N$^2$LO truncation errors (at some low confidence level) and
appears to show a clear tendency of increasing with growing values of $A$. Extending the
calculations to N$^3$LO will allow us  in the future to perform a more reliable and
elaborate estimation of the truncation uncertainty.

\begin{figure}[tb]
  \includegraphics[width=0.85\columnwidth]{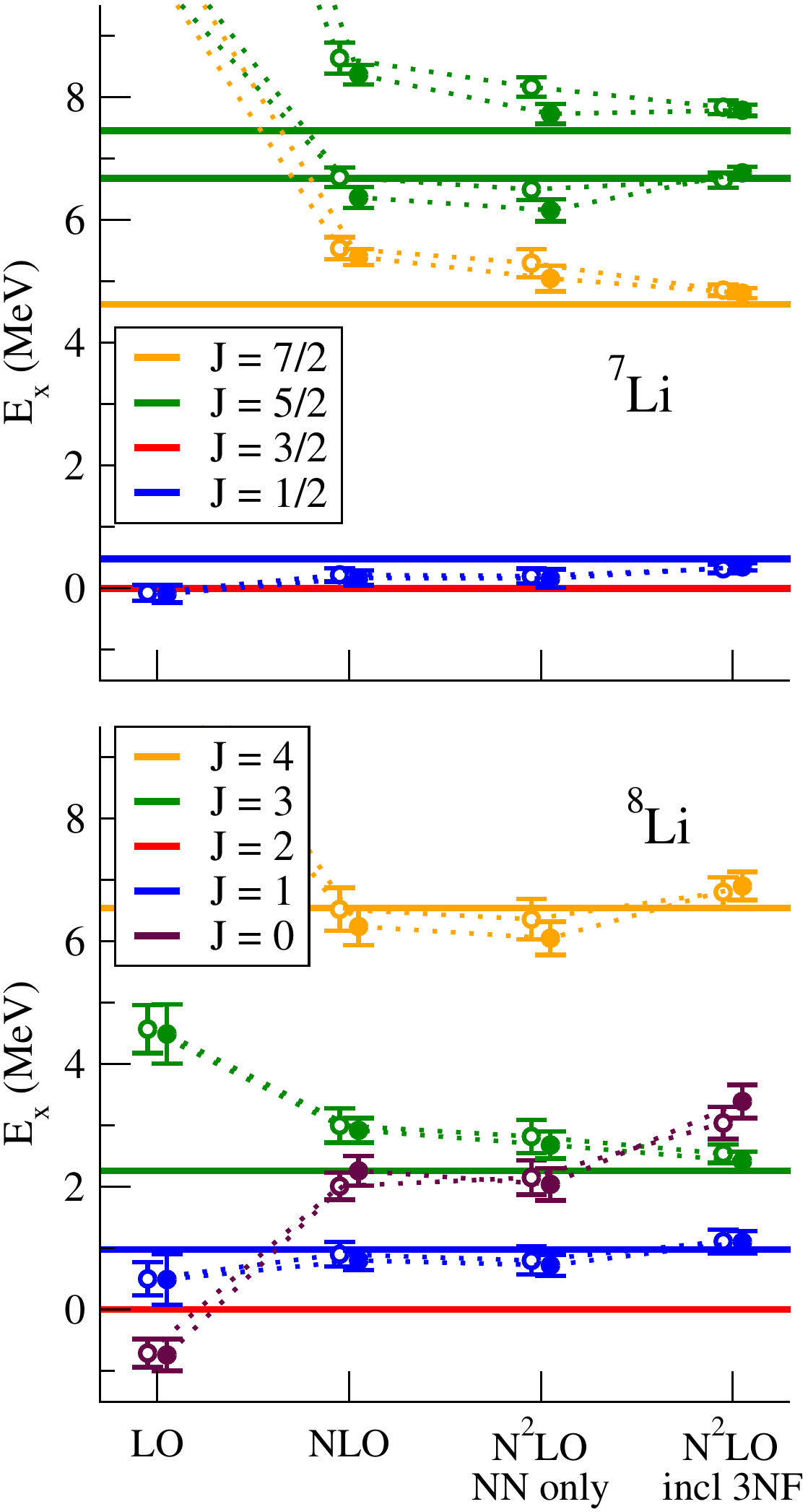}
  \caption{\label{Fig:res_spectra_7Li8Li}
    (Color online) Order-by-order excitation spectra of $^7$Li (top) and $^8$Li (bottom).
    All excitation energies are obtained with SRG parameter
    $\alpha=0.08$~fm$^4$; open symbols are with $\Lambda=450$~MeV,
    closed symbols are with $\Lambda=500$~MeV, and horizontal lines
    indicate experimental values~\cite{TILLEY20023,TILLEY2004155} .}
\end{figure}

In Table~\ref{Tab:res_A7A8A10A12} we also list our order-by-order
results for the excitation energies.  Again, these excitation energies
are the difference of the extrapolated total energies.  For $^7$Li our
results are, starting from NLO, in good qualitative agreement with
experiment, see Fig.~\ref{Fig:res_spectra_7Li8Li}.  
%%%%%%%%%%%%%%%%%%%%%%%%%%%%%%%%%%%%%%%%%%%%%%%%%%%%%%%%%%%%%%%%%%%%%%%%%%%%%
%%%%%%%%%%%%%%%%%%%%%%%%%%%%%%%%%%%%%%%%%%%%%%%%%%%%%%%%%%%%%%%%%%%%%%%%%%%%
However, at LO the spectrum looks very different: the
ground state and first excited state are nearly degenerate, and
reversed in order, while the second and third excited state are
significantly too high and also nearly degenerate.  This can easily be
qualitatively explained in terms of the clustering: the lowest two
states in $^7$Li can be viewed as a bound states of $^3$H and $^4$He
in an $L=1$ state with the spin and orbital motion (anti)aligned,
whereas the second and third excited state are bound states of $^3$H
and $^4$He in an $L=3$ state with the spin and orbital motion
(anti)aligned.  Without sufficient spin-orbit splitting in the
NN-potential at LO, the first two states become degenerate, as do the
second and third state.  Note that the second excited $\frac{5}{2}^-$
state has a different structure, and is even higher in the spectrum at
LO.  Starting from NLO however, the spectrum is in qualitative
agreement with experiment, and the differences between the excitation
energies between NLO and N$^2$LO (with or without 3NFs) are less than
an MeV; and at N$^2$LO (with 3NFs) there is good agreement with the 
experimental values.

Also for $^8$Li we see a qualitative difference between the spectrum at LO 
and at higher orders: at LO the ground state is actually a $0^+$ state, 
whereas the experimental ground state is a $2^+$ state.  
(Note that there is no narrow $0^+$ listed in Ref.~\cite{TILLEY2004155}.)  Starting from NLO,
the excitation energies of the $1^+$, $3^+$, and $4^+$ states are in reasonable 
agreement with the known experimental values, with only small changes as one goes 
from NLO to N$^2$LO without and with 3NFs; the latter gives best agreement for 
the low-lying spectrum.  In addition to the known narrow $1^+$, $3^+$, and $4^+$ states, 
and the $0^+$ state which is the ground state at LO, we also see evidence for 
additional $1^+$ and $2^+$ states between 3 and 7 MeV at N$^2$LO with 3NFs.  
These states are also found with the SCS interactions~\cite{Maris:2019llm};
however are not very well converged and probably correspond to broad resonances.

%%%%%%%%%%%%%%%%%%%%%%%%%%%%%%%%%%%%%%%%%%%%%%%%%%%%%%%%%%%%%%%%%%%%%%%%%%%%%
\begin{figure*}[tb]
  \includegraphics[width=0.9\textwidth]{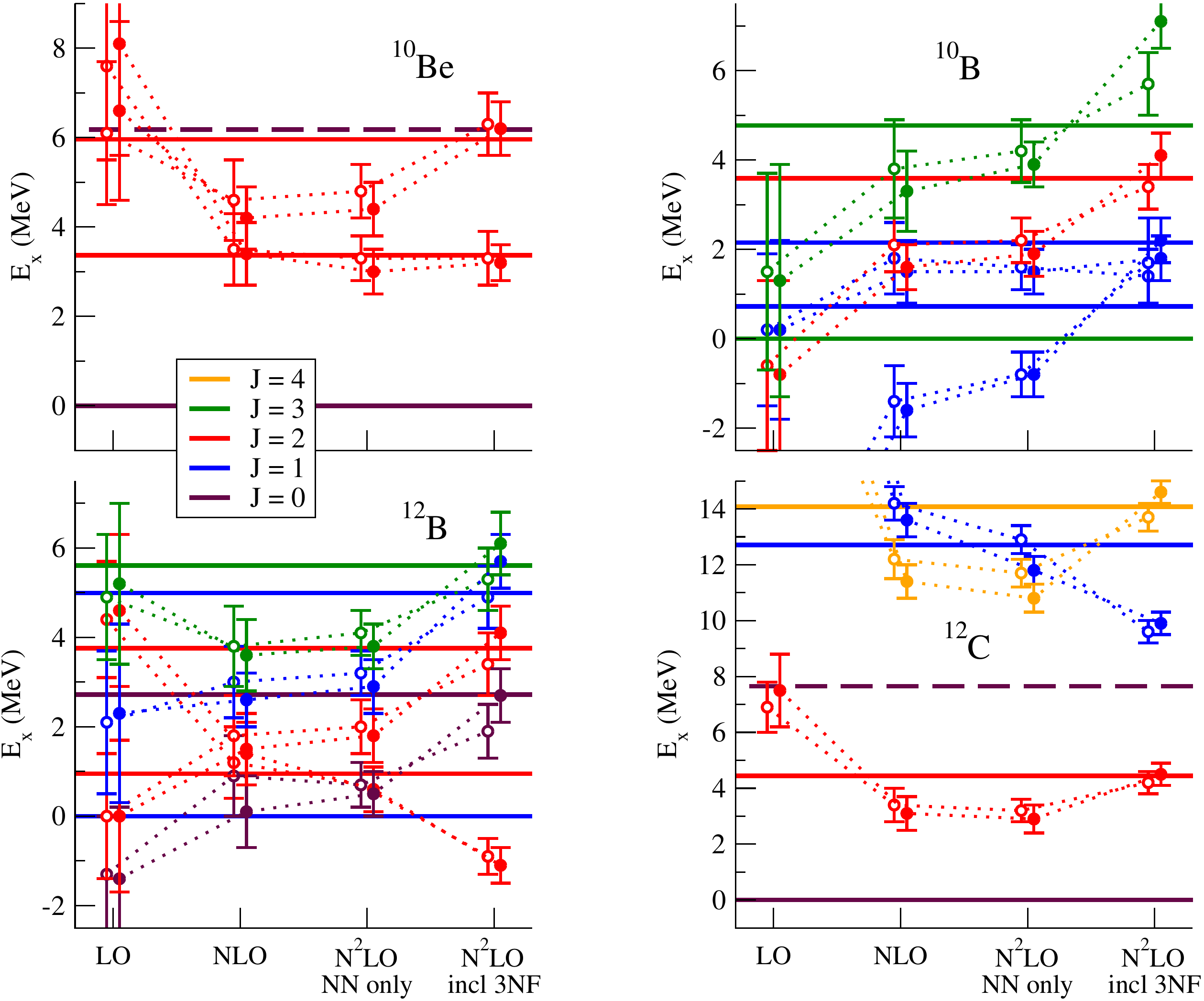}
  \caption{\label{Fig:res_spectra_A10A12}
    (Color online) Order-by-order excitation spectra of $^{10}$Be (top-left),
    $^{10}$B (top-right), $^{12}$B (bottom-left), and $^{12}$C (bottom-right).
    All excitation energies are obtained with SRG parameter
    $\alpha=0.08$~fm$^4$; open symbols are with $\Lambda=450$~MeV,
    closed symbols are with $\Lambda=500$~MeV, and horizontal lines
    indicate experimental values~\cite{TILLEY2004155,KELLEY201771} ; dashed lines indicate $0^+$ states not
    in the low-lying spectra  of our NCCI calculations.}
\end{figure*}
%%%%%%%%%%%%%%%%%%%%%%%%%%%%%%%%%%%%%%%%%%%%%%%%%%%%%%%%%%%%%%%%%%%%%%%%%%%%

The two lowest excited states of $^{10}$Be are both $J=2$ states, and at 
N$^2$LO their excitation energies are in good agreement with the experimental values, 
see Fig~\ref{Fig:res_spectra_A10A12}.  Although they have the same quantum numbers, 
they are easily distinguished by their quadrupole moments: at N$^2$LO the first 
$J=2$ state has a negative quadrupole moment and can be identified as a rotational
excitation of the ground state~\cite{Maris:2014jha}, whereas the second $J=2$ has 
a positive quadrupole moment.  Note however that at LO the order of these two states 
is reversed. In addition to these two $J=2$ states, there is also a narrow $J=0$ excited 
state at $6.179$~MeV (see Ref.~\cite{TILLEY2004155}, for which we do not find any evidence in our NCCI calculations.
It is unclear whether that is a deficiency of the NCCI approach or a property of the
interaction; also with the SCS interactions we do not see this excited 
$J=0$ state in the low-lying spectrum~\cite{Epelbaum:2018ogq,Maris:2019llm}.

The low-lying spectrum of $^{10}$B is known to be very sensitive to details of the 
interaction, and in particular to 3NFs~\cite{Navratil:2007we}.  Indeed, in Fig.~\ref{Fig:res_spectra_A10A12} 
we see that without 3NFs we do not get the correct ground state: one of the two low-lying 
$J=1$ states is (well) below the lowest $J=3$ state at LO, NLO, and N$^2$LO without 3NFs.  
Only after adding the 3NFs at N$^2$LO do we get the correct ground state.  A complicating
factor is that the two low-lying $J=1$ states mix as function of the basis parameters 
\hw\ and \nmax, which makes the extrapolation to the complete basis less reliable. 
Nevertheless, at N$^2$LO with 3NFs we find reasonable agreement for the low-lying 
spectrum, given our numerical uncertainty estimates.  Furthermore,
it is interesting to note that the convergence pattern with chiral order, and the effect
of the 3NFs, is very similar for the lowest $J=1$ state, the $J=2$ state, and the first 
excited $J=3$ state, indicating strong correlations between these three states.  

The spectrum of $^{12}$B turns out to be even more sensitive to the NN and 3N interaction,
as can be seen in the lower-left panel of Fig.~\ref{Fig:res_spectra_A10A12}.  There are two narrow $J=2$ excited states in $^{12}$B, which have an opposite behavior as we go from LO to N$^2$LO in the chiral expansion: one of them is almost degenerate with the $J=1$ ground state
at LO, but its excitation energy increases at NLO, and increases further at N$^2$LO (including 3NFs), whereas the excitation energy of the second $J=2$ state at LO decreases
at higher orders, and this $J=2$ state drops below the physical $J=1$ ground state
at N$^2$LO with 3NFs, becoming (incorrectly) the predicted ground state.  
Also note that there are three excited states, with $J=0$, $1$, and $2$, whose excitation
energies have a very similar pattern as function of the chiral order -- and this pattern is also very similar to that seen for three low-lying states
in $^{10}$B.  This suggests that these states have a closely related structure, which deserves further investigation.  It also remains to be seen what happens at higher chiral orders.

Last but not least, the spectrum of $^{12}$C.  As already mentioned, the Hoyle state 
(the first excited $J=0$ state in $^{12}$C, which is near the $3\alpha$ threshold) cannot 
be correctly described within the NCCI approach within the numerically accessible
basis space.  In Fig.~\ref{Fig:res_spectra_A10A12} this state is represented by the dashed
line.  The $J=2$ and $J=4$ states are rotational excitations of the ground state~\cite{Dytrych:2016vjy},
and it is therefore no surprise that their behavior as a function of the chiral order
is very similar; the ratio of their respective excitation energies remains nearly constant
at about 3.3, as one would expect for a rotational band.  More interesting and puzzling 
is the systematic decrease with chiral order of the $J=1$ (with $T=0$) state, 
in particular with the inclusion of the 3NFs at N$^2$LO.  
Again, the question is what happens with this state at higher chiral orders.

To summarize, most of the calculated spectra of $p$-shell nuclei show good agreement 
with experiment for the lowest narrow states with natural parity, with only a few exceptions.  
These exceptions are the $J=0^+$ excited states in $^{10}$Be and $^{12}$C (which are most
likely absent in our calculations due to the limited basis spaces), as well as several
states that are particularly sensitive to the details of the NN-potential and the 3NFs.
Specifically, these are the excited $J=1$ state in $^{12}$C;
the $J = 2$ state in $^{12}$B, which is the second excited $J=2$ state at
LO and NLO, but which becomes the theoretical ground state at N$^2$LO, 
whereas the experimental ground state has $J=1$; and the lowest states in $^{10}$B,
which has experimentally a $J=3$ ground state, with two low-lying $J=1$ excited states, 
but in ab initio calculations without 3NFs, typically one of these $J=1$ states becomes the lowest state.
However, in order to judge whether or not these exceptions are problematic, we need to consider
not only the extrapolation uncertainties, but also the chiral truncation uncertainties.

%%%%%%%%%%%%%%%%%%%%%%%%%%%%%%%%%%%%%%%%%%%%%%%%%%%%%%%%%%%%%%%%%%%%%%%%%%%%%
\section{Correlated truncation uncertainties for nuclear spectra} \label{Sec:correlated_spectra}
In this section we consider the EFT truncation errors  for the calculated  spectra summarized in Tables~\ref{Tab:res_A4A6} and \ref{Tab:res_A7A8A10A12}.
As described in Sec.~\ref{sec:Nd_scattering}, these uncertainties can be estimated using a Bayesian statistical model that learns from the order-by-order convergence pattern.
This model has been applied in Secs.~\ref{sec:Nd_scattering} and \ref{Sec:few-body_nuclei} in a \emph{pointwise} form, meaning that different observables are treated as statistically independent.
If applied to the energy spectra from the last section, one would add individual errors in quadrature to find the error bars for excitation energies (because they are a difference between excited- and ground-state energies treated independently). 
But as already noted and from all other experience, these excitation energies are generally much better determined than energies of the individual levels.
Therefore, to avoid overestimating the truncation errors it is essential to apply a \emph{correlated} error model, which we do here.

An extended model applicable to correlated truncation errors was recently developed in Ref.~\cite{Melendez:2019izc} and applied to infinite matter in Refs.~\cite{Drischler:2020hwi,Drischler:2020yad}. 
This model employed Gaussian processes (GPs) because the observables were continuous functions of the input variables, namely energy and density, respectively.
The spectra here are discrete, but we can adapt the GP results because every finite number of inputs will have a joint Gaussian distribution.
Rather than learn the hyperparameters of a covariance function that depends on the continuous distance between inputs, we can learn the covariance structure between discrete energy levels and nuclei from the observed pattern of order-by-order expansion coefficients $c_i$ [defined in Eq.~\eqref{eq:X_expansion}].

%%%%%%%%%%%%%%%%%%%%%%%%%%%%%%%%%%%%%%%%%%%%%%%%%%%%%%%%%%%%%%%%%%%%%%%%%%%%%
\begin{figure}[tb]
  \includegraphics[width=\columnwidth]{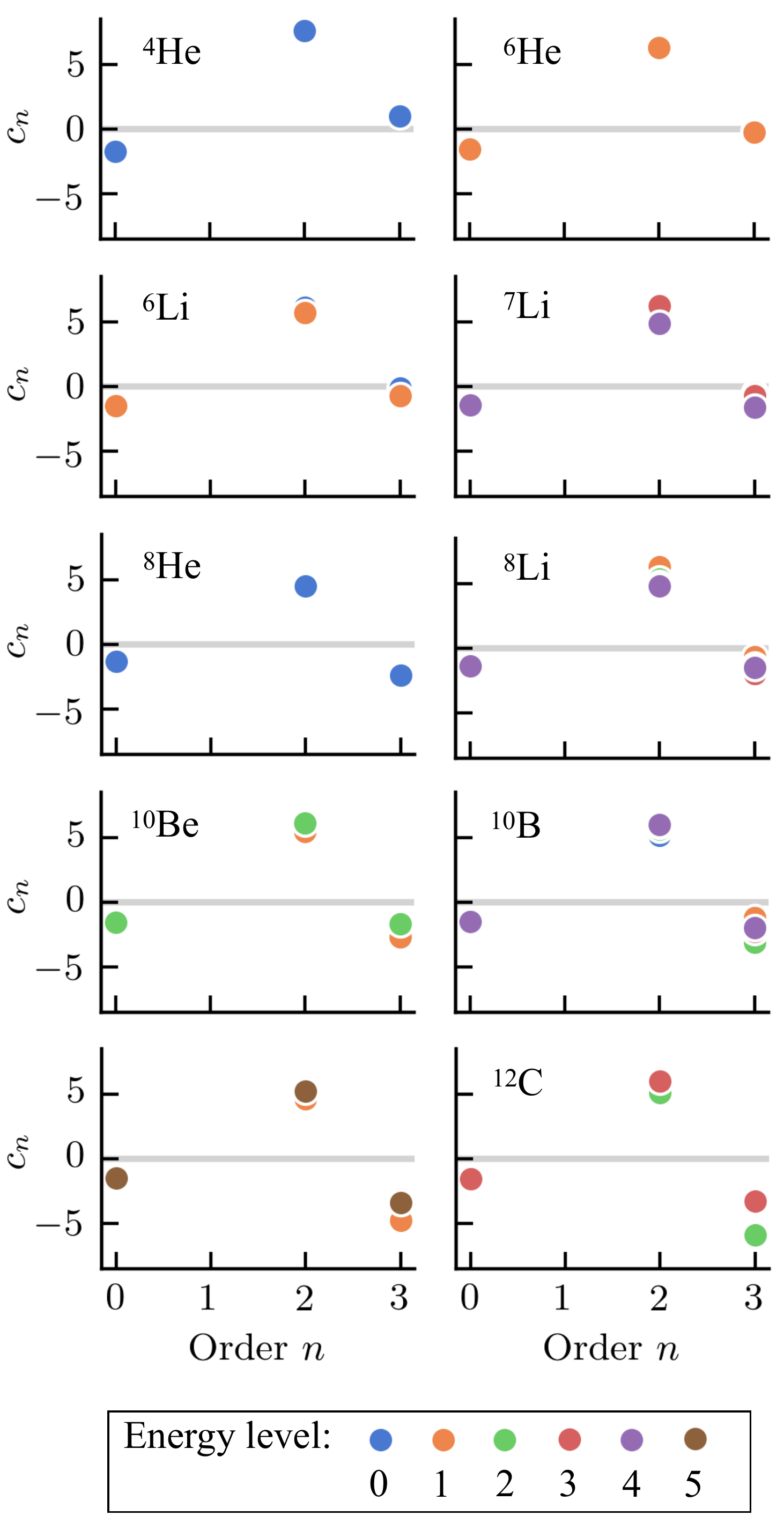}
  \caption{\label{Fig:spectra_coefficients}
    (Color online) Expansion coefficients for the individual energy levels in Tables~\ref{Tab:res_A4A6} and \ref{Tab:res_A7A8A10A12} with $\alpha = 0.08$~fm$^4$ and $\Lambda=450$~MeV. These are extracted according to Eq.~\eqref{eq:X_expansion} with a fixed value of $Q \approx 0.31$ and $X_{\rm ref}$ taken from experiment~\cite{TILLEY20023,TILLEY2004155,KELLEY201771} (or the N$^2$LO result for the $0^+$ in $^8$Li).
    }
\end{figure}
%%%%%%%%%%%%%%%%%%%%%%%%%%%%%%%%%%%%%%%%%%%%%%%%%%%%%%%%%%%%%%%%%%%%%%%%%%%%

To manifest the correlations, we plot the $c_i$ coefficients for each individual level listed in Tables~\ref{Tab:res_A4A6} and \ref{Tab:res_A7A8A10A12} in Fig.~\ref{Fig:spectra_coefficients}.
For this rough visualization, we extract the $c_i$s using a fixed $Q = M_\pi^{\rm  eff} / \Lambda_b = 200 / 650 \approx 0.31$ for all the states, with $X_{\rm ref}$ taken from experiment (or the N$^2$LO result for the $0^+$ state in $^8$Li).
We see high correlation as expected between observable coefficients for the spectra of a given nucleus but also between nuclei.
To model these correlations, we introduce a covariance matrix and determine it empirically~\cite{scikit-learn}.
We emphasize that the correlations shown beyond $c_0$ are for the \emph{corrections} to the observables.

%%%%%%%%%%%%%%%%%%%%%%%%%%%%%%%%%%%%%%%%%%%%%%%%%%%%%%%%%%%%%%%%%%%%%%%%%%%%%
\begin{figure}[!tb]
  \includegraphics[width=\columnwidth]{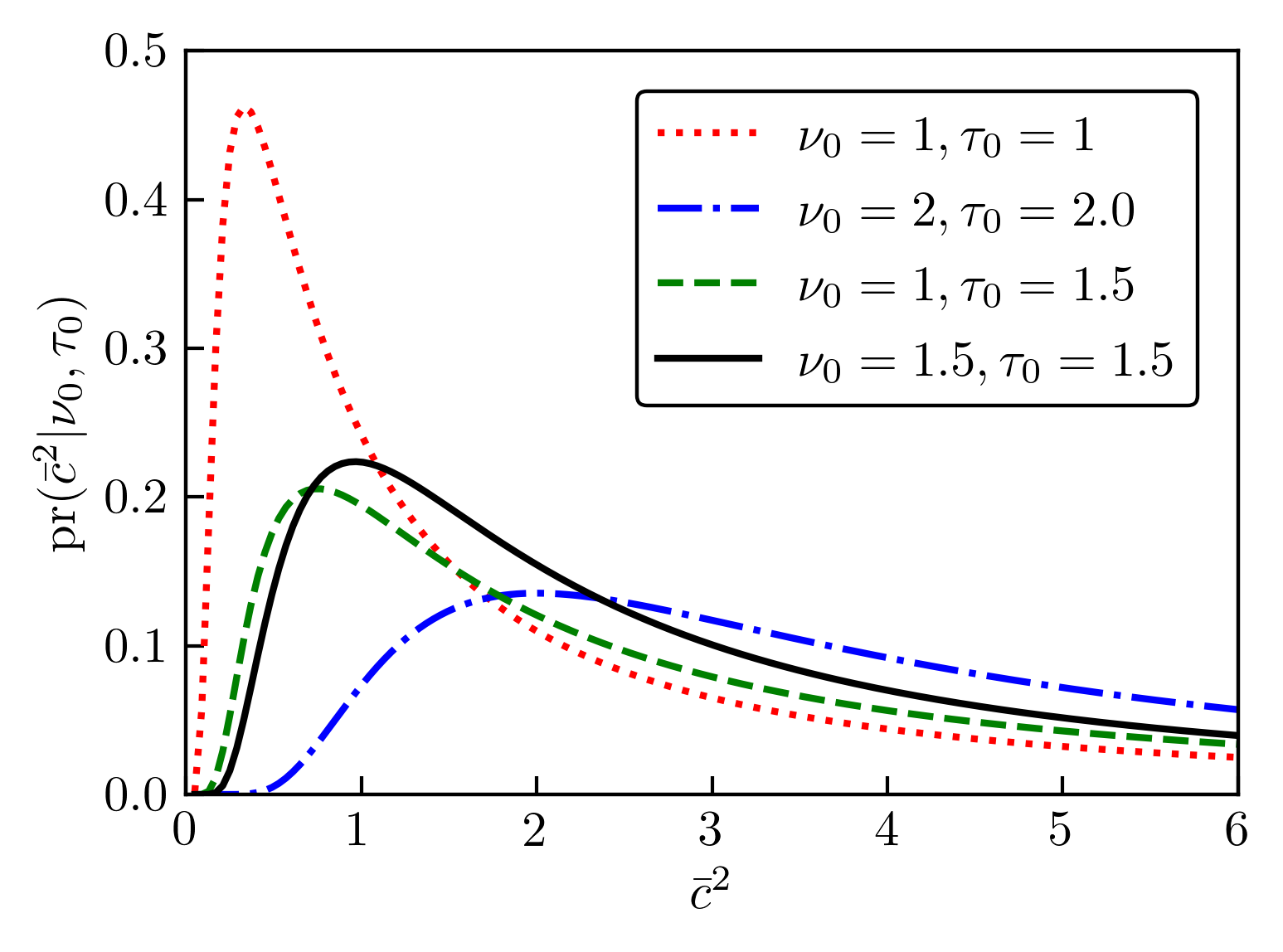}
  \caption{\label{Fig:prior_for_cbarsq}
    (Color online) Prior pdf for the variance $\cbar^2$ of the expansion coefficient with several choices of hyperparameters $\nu_0$ and $\tau_0$.}
\end{figure}
%%%%%%%%%%%%%%%%%%%%%%%%%%%%%%%%%%%%%%%%%%%%%%%%%%%%%%%%%%%%%%%%%%%%%%%%%%%%

As already seen in Eqs.~\eqref{eq:X_expansion} and \eqref{eq:ci_prior}, the truncation error model is contingent on the expansion parameter $Q$ and the characteristic variance $\cbar^2$ of the observable expansion coefficients $c_i$.
Ideally we would learn $Q^2$ and $\cbar^2$ from the order-by-order calculations together with the prior expectations for each.
A complication for the spectra of light nuclei is that the order-by-order convergence pattern is obscured for those observables at low orders by the strong cancellation between kinetic and potential energies~\cite{Binder:2018pgl}.
This is exacerbated in the present case by only having orders up to N$^2$LO. 

As a first approach we bypass the problem with the low orders by using just the $c_3$ coefficients to learn $\cbar^2$.
By ``learning'' we mean obtaining a statistical solution to the inverse problem of determining the distribution the coefficients come from (which is characterized by $\cbar^2$).
We follow Appendix~A of Ref.~\cite{Melendez:2019izc} and use a hierarchical model that is computationally efficient and enables us to both parameterize our prior expectations and easily marginalize (i.e., integrate over) the hyperparameters to reduce sensitivity.
Previous work has shown little sensitivity to the choice of prior once higher orders are available.
With only up to N$^2$LO available we can expect more sensitivity here, but this will be cured in future work with N$^3$LO and higher, so we do not exhaustively test the dependence on the choice of prior. 
For the analysis of spectra we use the scaled inverse-$\chi^2$ conjugate prior proposed in Ref.~\cite{Melendez:2019izc}, which is shown in Fig.~\ref{Fig:prior_for_cbarsq} for several candidate choices of the hyperparameters $\nu_0$ and $\tau_0$ to assess sensitivity.
Based on these tests we have chosen $\nu_0 = 1.5$, $\tau_0 = 1.5$ for the present error analysis.

If we use just the $c_3$ coefficients but from all of the energy levels to find a posterior for $Q$ (without accounting for correlations), it peaks close to the value adopted for Fig.~\ref{Fig:spectra_coefficients} ($Q\approx 0.3$), which is an average of the somewhat smaller values found for the lighter nuclei and the somewhat larger values found for the heavier nuclei.
This is consistent with expectations that $Q$ should increase with the increasing average kinetic energy (the use of the non-observable kinetic energy in estimating $Q$ is discussed in Ref.~\cite{Binder:2018pgl}).
To fit the empirical covariance matrix, it is not sufficient to use the $c_3$ results.
As a start, we also include $c_2$ for determining the covariance, which might overestimate the degree of correlation based on a comparison of orders in Fig.~\ref{Fig:spectra_coefficients}.
Other strategies for determining correlations for energy spectra will be explored in future work.

%%%%%%%%%%%%%%%%%%%%%%%%%%%%%%%%%%%%%%%%%%%%%%%%%%%%%%%%%%%%%%%%%%%%%%%%%%%%%
\begin{figure}[!tb]
  \includegraphics[width=\columnwidth]{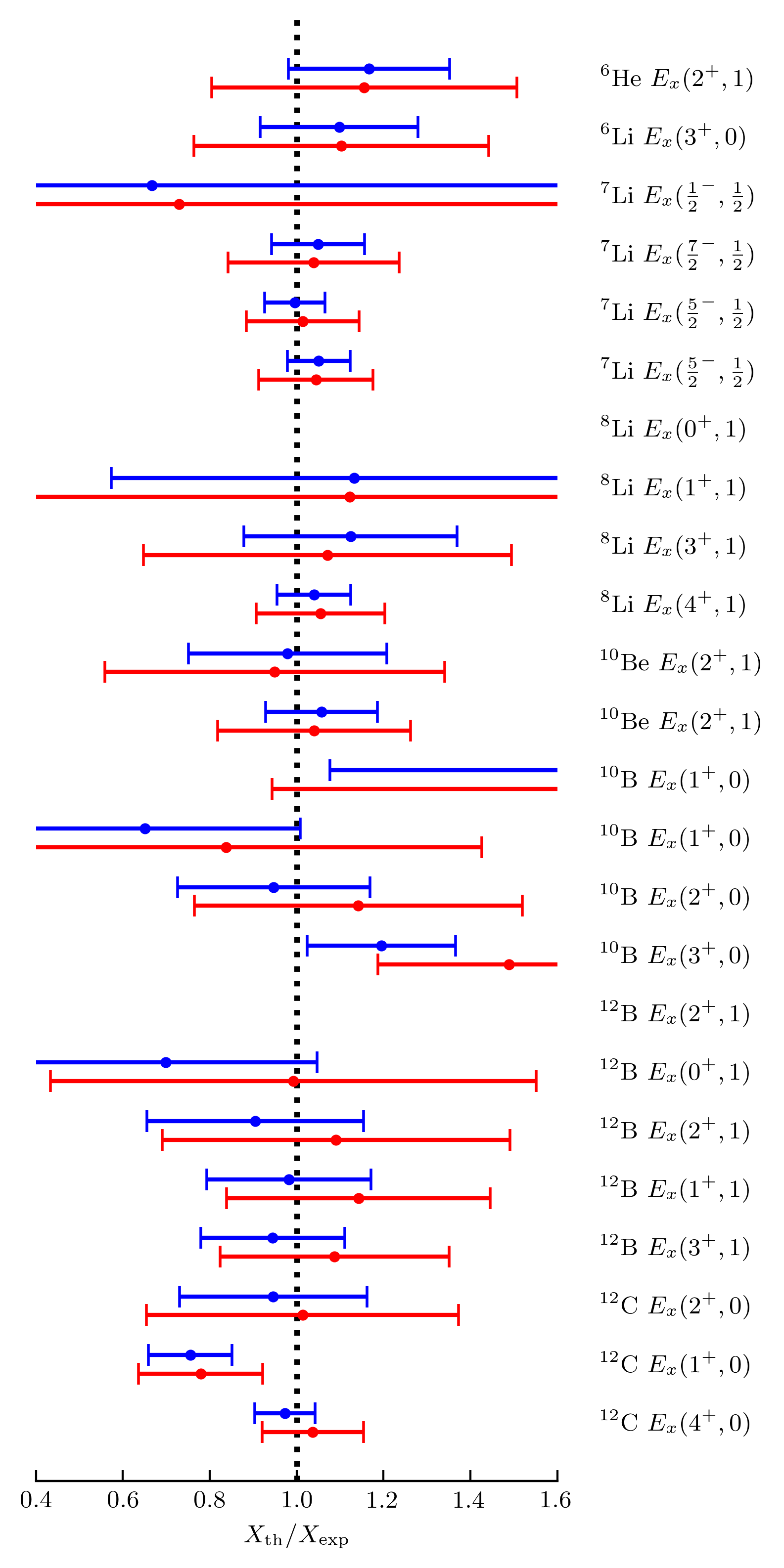}
  \caption{\label{Fig:excitation_energies_with_truncation_errors}
    (Color online) Central values (dots) for the excitation energies from Tables~\ref{Tab:res_A4A6} and \ref{Tab:res_A7A8A10A12} with 95\% Bayesian confidence intervals for truncation errors only indicated as error bars. 
    (We omit the $0^+$ in $^8$Li because an experimental value is not available and the lowest $2^+$ in $^{12}$B because the experimental ground state is not correctly predicted.)
    For each excitation from the calculated ground state, the upper (blue) bar is for $\Lambda = 450\,$MeV and the lower (red) bar is for $\Lambda=500\,$MeV. 
    All results shown are for $\alpha=0.08\,\mbox{fm}^{4}$. 
    }
\end{figure}
%%%%%%%%%%%%%%%%%%%%%%%%%%%%%%%%%%%%%%%%%%%%%%%%%%%%%%%%%%%%%%%%%%%%%%%%%%%%

The resulting Bayesian 95\% confidence intervals for the excitation energies are shown in Fig.~\ref{Fig:excitation_energies_with_truncation_errors}, where we plot the ratio of theory to experiment. 
The intervals are shown for the results from Tables~\ref{Tab:res_A4A6} and \ref{Tab:res_A7A8A10A12} for both the $\Lambda = 450\,$MeV (upper, blue) and  $\Lambda=500\,$MeV (lower, red) potentials at SRG $\alpha = 0.08\,\mbox{fm}^{4}$.
The truncation uncertainties are particularly large for very low-lying excitations in this representation because of the small numbers involved.
We see that the  uncertainties for the 500\,MeV potential are systematically larger than those for 450\,MeV potential, but in both cases the empirical coverage of experiment is good.
That is, the error bars encompass unity at roughly the rate one would expect for 95\% intervals.
We emphasize that without taking correlations into account, the intervals would have been significantly larger, and therefore would have been too conservative based on this comparison (i.e., with poor empirical coverage).

When N$^3$LO results are available, we will be able to validate these results and explore the covariance structure in greater detail.
It will be interesting to analyze the correlations among states with similar and distinct characteristics as expected from theoretical considerations.
We will also seek to make use of the lower-order results along the lines discussed in Ref.~\cite{Binder:2018pgl} as alternative approaches to the truncation errors.

%%%%%%%%%%%%%%%%%%%%%%%%%%%%%%%%%%%%%%%%%%%%%%%%%%%%%%%%%%%%%%%%%%%%%%%%%%%%%
\section{Summary and conclusions}
\label{conclusion}
In this paper we have, for the first time, applied the novel SMS
chiral NN potentials of Ref.~\cite{Reinert:2017usi}, along with the consistently
regularized 3NFs comprising subtraction terms, to study selected observables in Nd elastic
scattering and the deuteron breakup process as well as various properties of
light nuclei up to N$^2$LO in chiral EFT.  Our main findings can be
summarized as follows:
\begin{itemize}
\item
 We have used the approach introduced and advocated in Ref.~\cite{Epelbaum:2018ogq} to
 determine the LECs $c_D$ and $c_E$ entering the leading 3NF using the
 triton binding energy and the Nd cross section minimum at $E_N =
 70$~MeV for the cutoff values of $\Lambda = 450$ and $500$~MeV.
The resulting values of these LECs are found to be of natural size. 
The predicted results for Nd scattering observables sensitive to the
3NF that have been considered in Ref.~\cite{Epelbaum:2018ogq}, namely
the doublet scattering length $^2a$, Nd total cross section and the
differential cross section in elastic Nd scattering in the minimum
region at $E_N = 108$ and $135$~MeV, are found to be in
agreement with the experimental data from
Refs.~\cite{Abfalterer:2001gw,Sekiguchi:2002sf,Ermisch:2005kf}.
The only exception are the data of Ref.~\cite{Sekiguchi:2002sf}
at $E_N = 135$~MeV, for which a slight discrepancy (at the level of
$\sim 1.5 \sigma$ is observed). Further, in line with earlier studies, our
results suggest that $^2a$ is not suitable for the determination of
these LECs when used in combination with the $^3$H binding energy due
to the well-known strong correlation between these two observables. Our
predictions for the considered analyzing powers in elastic Nd
scattering at $E_N = 70$~MeV as well as for the differential cross
section and nucleon vector analyzing power $A_y$ in selected
deuteron breakup configurations at $E_N = 65$~MeV are in 
reasonable agreement with the data. 
We demonstrated the agreement between complete N$^2$LO predictions 
based on the SCS force~\cite{Epelbaum:2018ogq},
a previous form of the SMS interaction~\cite{Epelbaum:2019zqc}, 
and on the newly used SMS potential comprising subtraction terms in the 3NFs.

\item
For the ground state energies of light nuclei, we observe a similar
pattern to the one reported in our earlier study~\cite{Epelbaum:2018ogq} 
using the SCS chiral interactions at the
same chiral order. In particular, while the predicted binding
energies are found to be within $\sim 3\%$ of the experimental values for light
nuclei, a systematic overbinding trend sets in for $A \sim 9-10$ and
increases with growing $A$.  For the considered nuclei up to $A=12$,
our predictions are consistent with the experimental values within
errors. For the
lightest nuclei with $A=3,4$ we have also calculated the point-proton
and point-neutron radii. Our predictions for the point-proton
structure radii for $^3$H and $^3$He agree with the data within
errors.  For $^4$He, our N$^2$LO prediction for the radius is $\sim
4\%$ smaller than the central experimental value, but it is still
consistent with the datum at the $95\%$ confidence level. 
\item
We have addressed the question of quantifying truncation errors for
strongly correlated observables, such as the excitation energy
spectra, by using a correlated Bayesian error model and empirically
determining the corresponding covariance matrix. Our results for the
excitation energies are statistically consistent with both the assumed
expansion parameter and the experimental data for the spectra. 
\end{itemize}

In the future, we plan to extend these results in various
directions. First, it would be interesting to relax the constraint of
exactly reproducing the $^3$H binding energy employed in all our
calculations at N$^2$LO. This would require a more careful uncertainty
analysis in the determination of the LECs $c_D$, $c_E$ that would take
into account the expected truncation error for this observable. We
also plan to investigate the origin of the overbinding found for
heavier nuclei. In particular, it remains to be seen whether this
issue is related to deficiencies of the N$^2$LO approximation to the
NN force or has to be resolved by higher-order 3NF contributions.

Clearly, the most important step is the extension of these studies to
N$^3$LO, which will require the inclusion of the corresponding
corrections to the 3NF. However, it was shown in Ref.~\cite{Epelbaum:2019kcf} 
that one cannot apply the simple regularization approach we are
using in this study to the N$^3$LO contributions derived in
Refs.~\cite{Ishikawa:2007zz,Bernard:2007sp,Bernard:2011zr} using dimensional regularization, as this would 
destroy consistency with the NN interactions.
Rather, all loop contributions to the 3NF (and to exchange charge and
current operators) need to be re-derived using
a consistent semilocal regulator instead of the dimensional regularization. 
Work along these lines is in progress.

%%%%%%%%%%%%%%%%%%%%%%%%%%%%%%%%%%%%%%%%%%%%%%%%%%%%%%%%%%%%%%%%%%%%%%%%%%%%%%%%%
\section*{Acknowledgments}

This work was supported by BMBF (contract No.~05P18PCFP1 and 05P18RDFN1),
by the DFG SFB 1245 (Projektnummer 279384907), 
by the DFG and NSFC (Project-ID 196253076, TRR 110), 
by the VolkswagenStiftung (Grant No. 93562),
by the Polish National Science Center under Grants No. 2016/22/M/ST2/00173
and 2016/21/D/ST2/01120,
by the US National Science Foundation under Grant NSF PHY--1913069, 
and by the US Department of Energy under Grants DE-FG02-87ER40371, 
DE-SC0018223, DE-SC0018083, and DE-SC0015376.   
This research used resources of the National 
Energy Research Scientific Computing Center (NERSC) and
the Argonne Leadership Computing Facility (ALCF), which
are US Department of Energy Office of Science user facilities,
supported under Contracts No. DE-AC02-05CH11231 and
No. DE-AC02-06CH11357, and computing resources provided 
under the INCITE award `Nuclear Structure and Nuclear Reactions' from
the US Department of Energy, Office of Advanced Scientific Computing
Research. Further computing resources were provided on LICHTENBERG at 
the TU Darmstadt and on JURECA and the JURECA Booster 
of the J\"ulich Supercomputing Center, J\"ulich, Germany. 

%%%%%%%%%%%%%%%%%%%%%%%%%%%%%%%%%%%%%%%%%%%%%%%%%%%%%%%%%%%%%%%%%%%%%%%%%%%%%

%%%%%%%%%%%%%%%%%%%%%%%%%%%%%%%%%%%%%
\bibliographystyle{apsrev4-1}
\bibliography{lenpic_refs.bib}
%%%%%%%%%%%%%%%%%%%%%%%%%%%%%%%%%%%%%

%%%%%%%%%%%%%%%%%%%%%%%%%%%%%%%%%%%%%%%%%%%%%%%%%%%%%%%%%%%%%%%%%%%%%%%%%%%%%
\end{document}